\documentclass[acmtog, anonymous,nonacm]{acmart}

\usepackage{algpseudocode}
\usepackage{multirow}
\usepackage{microtype}
\usepackage{amsmath}
\usepackage{url}
\usepackage{caption}
\usepackage{graphicx}
\usepackage{mathrsfs}
\usepackage{booktabs}
\usepackage{bm}
\usepackage{subfigure}

\def\d{\ensuremath{\mathrm{d}}}

\usepackage[switch]{lineno}
\usepackage[ruled,linesnumbered]{algorithm2e}

\usepackage{amsmath}
\usepackage{xcolor}
\usepackage{ulem}
\usepackage{color}
\usepackage{epsfig}
\usepackage{multirow}
\newtheorem{theorem}{Theorem}

\newtheorem{lemma}[theorem]{Lemma}

\acmJournal{TOG}

\setcopyright{none}
\renewcommand\footnotetextcopyrightpermission[1]{} 
\begin{document}

\title{Anisotropic Gauss Reconstruction for Unoriented Point Clouds}

\begin{abstract}
	Unoriented surface reconstructions based on the Gauss formula have attracted much attention due to their elegant mathematical formulation and excellent performance. However, the isotropic characteristics of the formulation limit their capacity to leverage the anisotropic information within the point cloud. In this work, we propose a novel anisotropic formulation by introducing a convection term in the original Laplace operator. By choosing different velocity vectors, the anisotropic feature can be exploited to construct more effective linear equations. Moreover, an adaptive selection strategy is introduced for the velocity vector to further enhance the orientation and reconstruction performance of thin structures. Extensive experiments demonstrate that our method achieves state-of-the-art performance and manages various challenging situations, especially for models with thin structures or small holes. The source code will be released on GitHub.
	
\end{abstract}


\maketitle
\begin{figure*}[bpht]
	\centering
	\includegraphics[width=1\linewidth]{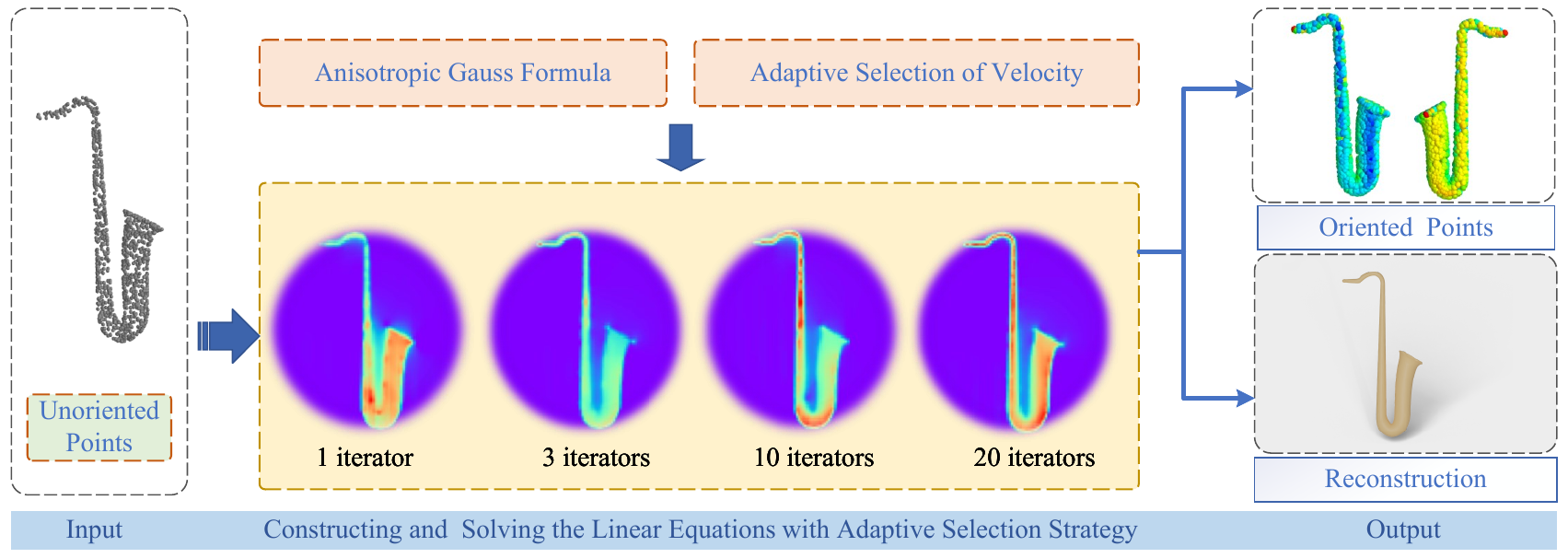}
	\caption{ We propose a novel anisotropic formulation by introducing a convection term in the original Laplace equation. By choosing different velocity vectors, we can fully take advantage of the anisotropic feature. Our method establishes and solves equation systems by the inputting unoriented points, simultaneously accomplishing orientation and reconstruction tasks. The color of the output point cloud in the top right-hand corner of the figure represents the normal information. Extensive experiments demonstrate that our method achieves state-of-the-art performance in both orientation and surface reconstruction for unoriented point clouds.}
	\label{fig:enter-label}
\end{figure*}
\section{Introduction}
Due to the convenience of point cloud acquisition, surface reconstruction plays a crucial role in computer graphics with a wide range of applications, such as geographic information systems, medical image processing, environmental modeling, and building visualization. 

In the past few decades, several well-established surface reconstruction methods have been proposed with remarkable performance.  ~\citet{kazhdan2006poisson,kazhdan2013screened} proposed converting the reconstruction problem into a spatial Poisson problem with additional constraints.
~\citet{manson2008streaming} chose orthogonal wavelet bases to calculate the indicator field.
~\citet{2019GR} introduce the Gauss formula to surface reconstruction and apply a modified kernel to address the near-singularity problem in computation. However, these methods have limitations in the requirement of oriented normals. In recent years, some attempts have been made for unoriented surface reconstruction. VIPSS ~\citep{huang2019variational} minimizes Duchon's energy using L-BFGS. Iterative Poisson surface reconstruction (iPSR) ~\citep{hou2022iterative}  introduces the idea of normal iteration. Parametric Gauss reconstruction (PGR) \cite{2022PGR}  constructs the linear system from the isotropic Gauss formula to calculate the linearized surface element. Converting unoriented point clouds into the oriented representation provides a novel approach to surface reconstruction. \citet{2023GCNO} proposed a smooth nonlinear objective function to characterize the requirements of an acceptable winding-number field, turn the problem into an unconstrained optimization problem, and reconstruct the surface by SPR.

Although there are many mature studies, surface reconstruction for unoriented points is still a challenging task. We notice that PGR gives us a good idea of reconstructing the surface by calculating the indicator function. However, PGR needs to solve under-determined equation systems and is sensitive to the regularization of equations. There is still room for improvement. This paper showcases our new research effort towards handling unoriented point clouds for both orientation and reconstruction. Considering that the isotropic Gauss formula will lead to the problem of lack of equations, we propose introducing a convection term in the original Laplace operator to obtain the anisotropic fundamental solution. Using the divergence theorem and double-layer potential theory, we derive the corresponding anisotropic Gauss formula to produce the indicator field. By choosing different velocity vectors as convection terms, the anisotropic Gauss formula constructs more linear equations to eliminate the dilemma of lack of equations. In addition, we provide two methods for solving under-determined and over-determined equations, respectively, and propose the blocking matrix strategy to save memory. Due to the introduction of anisotropy, our method reduces the sensitivity to the regularization of the equations. Our method can output the points' unit outward normals by normalizing the solution of the equations and can extract surfaces by the value of the indicator function on the query points with marching cubes. 

Moreover, the quality of velocity vectors can affect the performance of orientation and reconstruction. Therefore, we propose a novel adaptive selection strategy, especially for dealing with thin structures by principal component analysis (PCA) and singular value decomposition (SVD). We put forward detailed ideas for improving our method. We conduct thorough comparisons with other well-known methods through comprehensive experiments, which showcase our method's effectiveness and scalability on famous datasets, including thin structures, outliers, and noisy or sparse point clouds.

Our main contributions can be summarized as follows.
\begin{itemize}
	\item We introduce a convection term in the original Laplace operator to extend the Gauss formula into an anisotropic form. 
	\item By introducing the anisotropy, we construct more linear equations and decrease the sensitivity to regularization.
	\item We propose a novel adaptive selection strategy for velocity vectors, which can further improve orientation and reconstruction results, especially for point clouds with thin structures or small holes.
\end{itemize}
\section{Related Work}
\subsection{Surface Reconstruction based on Implicit Function }
The goal of implicit-based methods is to construct an implicit function or implicit field and make the input point cloud lie on one of its level sets. The task of surface reconstruction has been extensively researched in the past decades, and we classify these methods into two categories based on whether the normal information of the point cloud is needed. 

\subsubsection{ Surface Reconstruction for Oriented Point Clouds}

The radial basis function methods treat each point in the point cloud as the center of a radial basis function and adjust the weights of these functions to approximate the signed distance function (SDF). To solve larger scale point clouds problem,  ~\citet{923379}, \citet{walder2006implicit} proposed to use compactly supported radial bases. In order to further improve the results and utilize normal information, ~\citet{macedo2011hermite}, ~\citet{ijiri2013bilateral}, ~\citet{liu2016closed} propose to use Hermite RBFs to interpolate points' positions and normals together. Implicit moving least squares (IMLS) fit local algebraic shapes and minimize the total squared errors to nearby points and normals, such as \cite{dey2005adaptive}, \cite{kolluri2008provably}.

In addition to approximation and fitting, Stokes' theorem provides new inspiration. ~\cite{kazhdan2005reconstruction}, ~\cite{manson2008streaming}, ~\cite{ren2018biorthogonal}, and  ~\cite{2019GR} respectively choose Fourier bases, orthogonal wavelet bases, biorthogonal wavelet bases, isotropic fundamental solution of Laplace function, and compute the corresponding coefficient. \cite{kazhdan2005reconstruction} has a breakneck running speed at that time by using fast Fourier transforms (FFT). \citet{manson2008streaming}  utilizes the orthogonal compact support property of wavelet basis functions to maintain linear computational complexity and handle multi-scale reconstruction. This algorithm not only runs fast but also has the potential to handle large-scale point clouds. \citet{ren2018biorthogonal} further expands the selection space of wavelet bases. \citet{2019GR}  introduces the Gauss formula to surface reconstruction problems and has good reconstruction results.

Another popular choice is converting the reconstruction problem into a spatial Poisson problem, which has achieved good reconstruction results. This novel idea is first proposed in 
\cite{kazhdan2006poisson}. \cite{calakli2011ssd,kazhdan2013screened,kazhdan2020poisson} fit the smoothed indicator function near the target surface and enhance algorithm performance by adding different constraint information. 

\subsubsection{ Surface Reconstruction for Unoriented Point Clouds}
Attributed to the inherent challenge of acquiring normal information, numerous point clouds, particularly those derived from real-world data, encounter difficulties in being properly oriented. While addressing this issue is more complex and challenging, it has significant practical implications and potential applications. Much attention has been paid to such a problem.

Deep learning methods have been gradually introduced and thoroughly studied in surface reconstruction in recent years. These methods represent implicit fields with neural networks. Point2surf (P2S) \cite{erler2020points2surf} improves generalization performance and reconstruction accuracy by learning a prior over a combination of detailed local patches and coarse global information. \cite{gropp2020implicit} (IGR) proposes a new paradigm for computing high-fidelity implicit neural representations based on implicit geometric regularization. DiGS \cite{ben2022digs} incorporates soft second-order derivative constraints to guide the INR learning process, leading to better representations. Neural-Singular-Hessian (NSH) \cite{wang2023neural} enforces the Hessian of the neural implicit function to have a zero determinant for points near the surface, which suppresses ghost geometry and recovers details from unoriented point clouds with better expressiveness. However, these deep learning methods often require long-term training and solving time. Moreover, due to memory limitations, these methods often only handle sparse point clouds with weak robustness.

In recent years, some research has been proposed on handling unoriented point clouds based on traditional methods. These methods are primarily based on the mature oriented surface reconstruction methods with clever designs to eliminate the dependence on normals.
Shape as points (SAP) \citep{peng2021shape} revisits the classic point cloud representation and introduces a differentiable point-to-mesh layer by using a differentiable formulation of PSR ~\citep{kazhdan2006poisson}. Also based on PSR ~\citep{kazhdan2006poisson},  iterative Poisson surface reconstruction (iPSR) ~\citep{hou2022iterative} computes the normals from the surface in the preceding iteration and then generates a new surface with better quality. Based on GR \cite{2019GR}, parametric Gauss reconstruction (PGR)  \cite{2022PGR} takes unoriented point clouds as input and solves for a set of linearized surface elements that produce the indicator field represented by the Gauss formula.

\subsection{Normal Consistent Orientations for Point Clouds}

Calculating high-quality normals for unoriented point clouds is an important topic in geometric modeling and computer graphics. Converting unoriented point clouds into the oriented representation provides a novel approach to surface reconstruction. Overall, we can divide them into optimization-based and learning-based methods.

\textbf{Optimization-based Methods}  The research on optimization-based methods has a long history, which can date back to the last century. The pioneering work uses principal component analysis (PCA) to initialize normal orientations by ~\citet{hoppe1992surface}. Although many propagation-based methods have emerged, these methods always fail to handle complex or separated point clouds.

Over the years, many technologies have been proposed to mitigate the problems of relying solely on local information. Dipole ~\citep{metzer2021orienting} uses dipole propagation across patches iteratively. VIPSS ~\citep{huang2019variational} minimizes the Duchon's energy with the L-BFGS algorithm. ~\citet{xu2023globally} propose a smooth nonlinear objective function to characterize the requirements of an acceptable winding-number field ~\citep{mcintyre1993new} and turn the problem into an unconstrained optimization problem.  ~\citet{xiao2023point} propose combined iso-value constraints with local consistency requirements for normals to construct optimization formulation.

\textbf{Learning-based Methods}
Learning-based methods often treat oriented normal estimation as a classification or regression task where the normals are directly regressed from the feature extracted from the local patches. PCPNet ~\citep{guerrero2018pcpnet} encodes the multiple-scale features of local patches in a structured manner, which enables one to estimate local shape properties such as normals and curvature. ~\citet{li2023shs} propose a method for oriented normal estimation by aggregating the local and global information and learning signed hypersurfaces end-to-end.

Due to the tight link between orientation and surface reconstruction, some proposed state-of-the-art methods can simultaneously accomplish orientation and surface reconstruction tasks. For example, as reconstruction methods, iPSR ~\citep{hou2022iterative} and PGR \cite{2022PGR} both have the potential for orientation. The former can output the estimated normals from the estimated surface, while the latter can obtain the normals from the solution of the equation. However, iPSR  ~\citep{hou2022iterative} cannot be regarded as an orientation method strictly due to its resample of the input point cloud, especially for dense or noisy point clouds. Due to the lack of equations in PGR \cite{2022PGR}, there is room for further improvement in the orientation accuracy. This is also one of the motivations of our method.

\begin{figure}[htbp]
	\centering
	\includegraphics[width=1\linewidth]{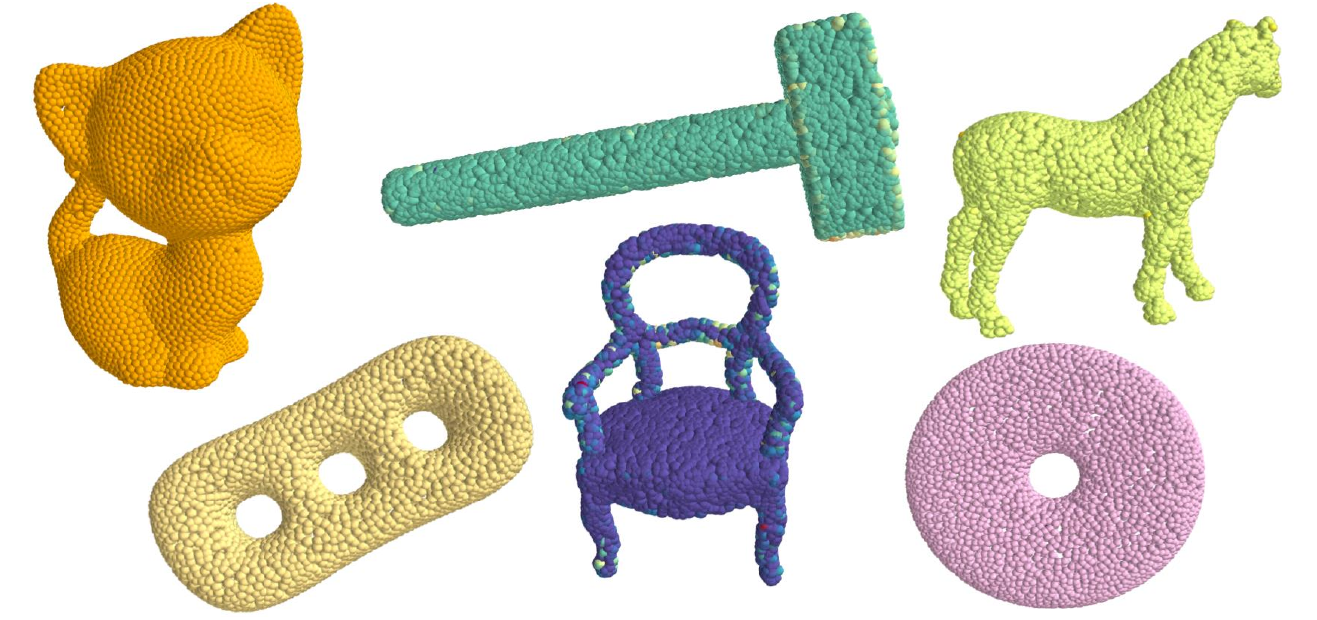}
	\caption{By introducing anisotropy, our method not only improves the quality of surface reconstruction but also stimulates the potential of the Gauss formula for orientation.}
	\label{fig:enter-label}
\end{figure}
\section{Motivation} \label{sec:pre}
An important and popular way to reconstruct the surfaces from the unoriented point clouds is to calculate the indicator function $\chi$. The Gauss formula provides a novel and powerful tool to represent the indicator function as a boundary integral.

In detail, let $\Omega \subset \mathbb{R}^{3}$ be an open and bounded region with the smooth boundary $\partial \Omega$. The indicator function $\chi(\bm{x})$ of $\Omega$ can be calculated through divergence theorem
\begin{align}\label{GL1}
	\int_{\partial \Omega}  \dfrac{\partial \Phi(\bm{x}-\bm{y})}{\partial \bm{n}(\bm{y})}\d S(\bm{y})&=\chi(\bm{x})=\left\{
	\begin{aligned}
		&0 & & {\bm{x} \in \mathbb{R}^3  \backslash \overline{\Omega}} \\
		&\dfrac{1}{2}  & &   {\bm{x} \in \partial \Omega} \\
		&1 & & {\bm{x} \in \Omega},
	\end{aligned} \right.\nonumber
\end{align}
where
\begin{align*}
	\Phi(\bm{x})=\dfrac{1}{4\pi|\bm{x}|}
\end{align*}
is the three-dimensional fundamental solution, which can be obtained by using symmetry to find its radial solution in Appendix \ref{A}. $\overline{\Omega}$ represents the closure of the region $\Omega$, $\bm{n}(\bm{y})$ represents the outward unit normal vector at any point $ \bm{y} \in \partial \Omega$ and $\mathrm{d} S(\bm{y})$ denotes the surface element of the  point $ \bm{y}$. $|\cdot|$ represents the $L_{2}$ norm of vectors. The proof will be presented as a special case of our generalized theorem later. The $n-$dimensional Gauss formula can be obtained by replacing $\Phi(\bm{x})$ with the $n-$dimensional Laplace fundamental solution in equation \eqref{E1}.

As shown in GR\cite{2019GR} and PGR\cite{2022PGR}, reconstructing the surfaces through the Gauss formula is feasible with good reconstruction results. However, PGR needs to solve under-determined equation systems. For $\mathcal{P}=\{\bm{p}_{i}\}_{i=1}^{N}$, the isotropic Gauss formula can only construct $N$ non-homogeneous equations, but there are $3N$ unknown variables. In addition, PGR is sensitive to the regularization of equations and shows poor performance under the inappropriate regularization value, which is shown in the supplementary material.

Extensive experiments show that PGR still has room for improvement in orientation, normal estimation, and surface reconstruction. On the one hand, the normals estimated by PGR deviate greatly from the ground truth. On the other hand, due to its reliance on isotropic fundamental solutions, PGR suffers from poor reconstructions for point clouds with thin structures or small holes. 

The isotropic Gauss formula utilizes the information of point clouds incompletely. It does not incorporate the geometric characteristics of the point cloud, especially the direction information. Due to the lack of equations and information, it is difficult to completely avoid the singularity of linear equations, leading to a strong dependence on regularization.

Inspired by this observation, we propose to stimulate the potential of the Gauss formula for reconstruction and orientation by introducing the anisotropy and constructing more equations to make full use of point cloud information. To generalize the fundamental solution, we add convection terms, including velocity vectors and first-order derivatives, into the original Laplace equation,
\begin{align}
	\Delta u -\bm{c}\cdot \nabla u = 0,
\end{align}\label{GL2}
where $\bm{c} \in \mathbb{R}^{3}$. 




We refer to the solution of equation \eqref{GL2} as the anisotropic fundamental solution, denoted as $\Phi_{c}$. we can derive an analytical solution
\begin{align}\label{GFF1}
	\Phi_{\bm{c}}(\bm{x})=\dfrac{1}{4\pi|\bm{x}|}e^{\frac{1}{2}(\bm{c}\cdot \bm{x} -|\bm{c}||\bm{x}|)},
\end{align}
with the gradient of it,
\begin{align}\label{GFF2}
	\nabla \Phi_{\bm{c}}(\bm{x})=\dfrac{1}{4\pi|\bm{x}|}e^{\frac{1}{2}(\bm{c}\cdot \bm{x}-|\bm{c}||\bm{x}|)}(-\dfrac{\bm{x}}{|\bm{x}|^{2}}+\dfrac{1}{2}\bm{c}-\dfrac{1}{2}|\bm{c}|\dfrac{\bm{x}}{|\bm{x}|}).
\end{align}

When $\bm{c}=(0,0,0)^{T}$, the anisotropic fundamental solution degenerates to be the isotropic solution $\Phi$. $|\cdot|$ represents the $L_{2}$ norm of vectors. \textbf{The complete and detailed theoretical derivation of the theorem is provided in Appendix \ref{B}.}

Using the divergence theorem and double layer potential theory, we can derive the anisotropic Gauss formula corresponding to \eqref{GFF1}.
\begin{theorem}[Anisotropic Gauss Formula]\label{THGGL}
	Let $\Omega \subset \mathbb{R}^{3}$ be an open and
	bounded region with the smooth boundary $\partial \Omega$, then the indicator function $\chi(\bm{x})$ of the region $\Omega$ can be calculated through the anisotropic fundamental solution \eqref{GFF1}. In detail,
	\begin{align}\label{GGL}
		\chi(\bm{x}) =\int_{\partial \Omega}K_{\bm{c}}(\bm{x}-\bm{y})\cdot \bm{n}(\bm{y}) \d S(\bm{y}) ,
	\end{align}
	where
	\begin{align}
		K_{c}(\bm{x})=\nabla \Phi_{\bm{c}}-\bm{c}\cdot \Phi_{\bm{c}}= \dfrac{1}{4\pi|\bm{x}|}e^{\frac{1}{2}(\bm{c}\cdot \bm{x}-|\bm{c}||\bm{x}|)}(-\dfrac{\bm{x}}{|\bm{x}|^{2}}-\dfrac{1}{2}\bm{c}-\dfrac{1}{2}|\bm{c}|\dfrac{\bm{x}}{|\bm{x}|}),
	\end{align}
	and $\bm{N}(\bm{y}) $ represents the outward unit normal vector at any point $ \bm{y} \in \partial \Omega$ and $\mathrm{d} S(\bm{y})$ denotes the surface element of the  point $ \bm{y}$. $|\cdot|$ represents the $L_{2}$ norm of vectors.
\end{theorem}

\textbf{The complete and detailed theoretical derivation of the Theorem \ref{THGGL}  is provided in Appendix \ref{C}.}
When $\bm{c}=(0,0,0)^{T}$, the anisotropic Gauss formula degenerates to be the isotropic Gauss formula.

\section{method}\label{sec:sec4}
Let the target area $\Omega \subset \mathbb{R}^{3}$ be an open and
bounded region with the smooth boundary $\partial \Omega$. The input point cloud is the points on the surface, $\mathcal{P}\in \partial \Omega$.

\subsection{Discretizing the Anisotropic Gauss Formula}
We first discretize the integral in equation \eqref{GGL}.
\begin{equation}\label{11}
	\begin{aligned}
		\chi(\bm{x}) & =\int_{\partial \Omega}K_{\bm{c}}(\bm{x}-\bm{y})\cdot \bm{n}(\bm{y}) \d S(\bm{y}) \\
		&\approx \sum_{j=1}^{N} \varphi_{j}^{\bm{c}}(\bm{x})\cdot \bm{n}_{\bm{p}_{j}} \sigma_{\bm{p}_{j}},
	\end{aligned}
\end{equation}
where
\begin{align*}\tiny
	\varphi^{\bm{c}}_{j}(\bm{x})=\dfrac{1}{8\pi|\bm{x}-\bm{p}_{j}|}e^{\frac{1}{2}(\bm{c} \cdot (\bm{x}-\bm{p}_{j})-|\bm{c}||\bm{x}-\bm{p}_{j}|)}(-\dfrac{2(\bm{x}-\bm{p}_{j})}{|\bm{x}-\bm{p}_{j}|^{2}}-\bm{c}-|\bm{c}|\dfrac{\bm{x}-\bm{p}_{j}}{|\bm{x}-\bm{p}_{j}|}),
\end{align*}
$\bm{n}_{\bm{p}_{j}}$ represents the outward unit normal vector at the point $ \bm{p}_{j}$, and $\sigma_{\bm{p}_{j}}$ denotes the surface element of the  point $ \bm{p}_{j}$. $|\cdot|$ represents the $L_{2}$ norm of vectors. Since we know the location information of the point cloud, $\varphi^{\bm{c}}_ {j}(\bm {x})=(\varphi^{\bm{c}}_{j,1}(\bm {x}),\varphi^{\bm{c}}_{j,2}(\bm {x}),\varphi^{\bm{c}}_{j,3}(\bm {x})) $ in the equation \eqref{11} are known variables, while the only unknown variables are the normal and area elements.

We denote the $3$-dimensional vector
\begin{equation}\label{LSE}
	\mu_{j}=(\mu_{j,1},\mu_{j,2},\mu_{j,3})  \triangleq \bm{n}_{\bm{p}_{j}} \sigma_{\bm{p}_{j}}
\end{equation}
to estimate area and normal information of $ \bm{p}_{j}$ and be called as \textbf{linearized surface element (LSE)}. Then, the anisotropic Gauss formula has a discrete form 
\begin{align}\label{EQ:D}
	\chi(\bm{x}) &\approx \sum_{j=1}^{N} \sum_{k=1}^{3}\varphi^{\bm{c}}_{j,k}(\bm{x})\mu_{j,k}
\end{align}
that can be used to calculate the indicator function of $\Omega$. Although, due to the lack of the outward normals, we cannot calculate the indicator function by estimating $\mu_{j,k}$ directly. However, we can still utilize the fact that the value of the indicator function at any point on the surface is $\frac{1}{2}$  to construct the equation systems. Then, we solve the equations to obtain the value of $\mu_{j,k}$.

\subsection{Constructing and Solving the Linear Equations}
For any fixed velocity vector $\bm{c}$, we select the input points $\bm{p}_{i},i=1,2,3,\cdots, N$ as the query point $\bm{x}$ in equation \eqref{EQ:D} sequentially, which lie on the surface of the region. Then, according to Theorem \ref{THGGL},
\begin{align*}
	\sum_{j=1}^{N} \sum_{k=1}^{3}\varphi^{\bm{c}}_{j,k}(\bm{p}_{i})\mu_{j,k}= \dfrac{1}{2},  \quad i=1,2,3,\cdots ,N, 
\end{align*}
and we denote it as
\begin{align}\label{overall}
	A^{\bm{c}}_{i}(\bm{p}_{i};\mathcal{P})\mu=\frac{1}{2},  \quad i=1,2,3,\cdots ,N
\end{align}
for brevity, where $\mu \in \mathbb{R}^{3N\times 1}$ denotes the flattened vector of $\mu_{j,k}$, and  $ A^{\bm{c}}_{i} \in \mathbb{R}^{1 \times 3N}$ is a row vector to represent the flattened vector of $\varphi^{\bm{c}}_{j,k}(\bm{p}_{i})$.
It should be pointed out that the selection of query points and velocity vectors is independent. We can choose different velocity vectors $\bm{c}_{\bm{p}_{i},s}$ for different point $\bm{p}_{i} $ based on the characteristics of such point, and there is no limit to the number $s$  of velocity vectors.

In order to reduce the number of hyperparameters, we select a fixed velocity vector $\bm{c}$ to generate $N$ equations. Namely
\begin{align}\label{111}
	A^{\bm{c}}(\mathcal{P};\mathcal{P})\mu=\frac{\bm{1}}{\bm{2}}.
\end{align}

However, when $\bm{x}=\bm{y}=\bm{p}_{j}$, the equation is singular and cannot be ignored.
The singularity may lead to an increase in error and a decrease in stability, orientation, and reconstruction. Hence, we modify the distance function $d(\bm{x},\bm{p}_{j})=|\bm{x}-\bm{p}_{j}|$ as 
\begin{align*}
	\tilde{d}(\bm{x},\bm{p}_{j})=\max\{d(\bm{x},\bm{p}_{j}),w(\bm{x})   \}, 
\end{align*}
where $w(\bm{x})$ is the width function to set the threshold of truncation. It is related to the average distance of the $k$ closest points to the point cloud and can be calculated as 
\begin{align*}
	w(\bm{x})=\max\{w_{\min},\sqrt{\dfrac{1}{k_{w}}\sum_{\bm{p}\in \text{kNN}(\bm{x};\mathcal{P})}|\bm{x}-\bm{p}|^{2}}\},
\end{align*}
where $\text{kNN}(\bm{x};\mathcal{P})$ represents the $k$ nearest neighbors of $\bm{x}$ in input point cloud $\mathcal{P}$, and $w_{\min}$, $k_{w}$ are hyperparameters. In most examples, $w_{\min}$ is suggested to be set as $0.0015$.
We denote the equation \eqref{111} after truncation as
\begin{align}
	A_{\bm{c}}(\mathcal{P};\mathcal{P})\mu=\frac{\bm{1}}{\bm{2}}.
\end{align}

We have completed the construction of $N$ equations induced by single velocity vector $\bm{c}$.
However, as shown in the PGR, the information provided by a single velocity vector is insufficient. Therefore, choosing multiple velocity vectors to construct more equations is recommended.

For $3$-dimensional point clouds $\mathcal{P}$ with $N$ points, the number of unknown variables is $3N$, but the number of equations varies with the selection of velocity vectors. Therefore, we provide two methods for solving under-determined and over-determined equations, respectively. When solving the square matrices, either approach can be used. We denote $m$ as the number of selected velocity vectors.

\begin{figure}
	\centering
	\includegraphics[width=1\linewidth]{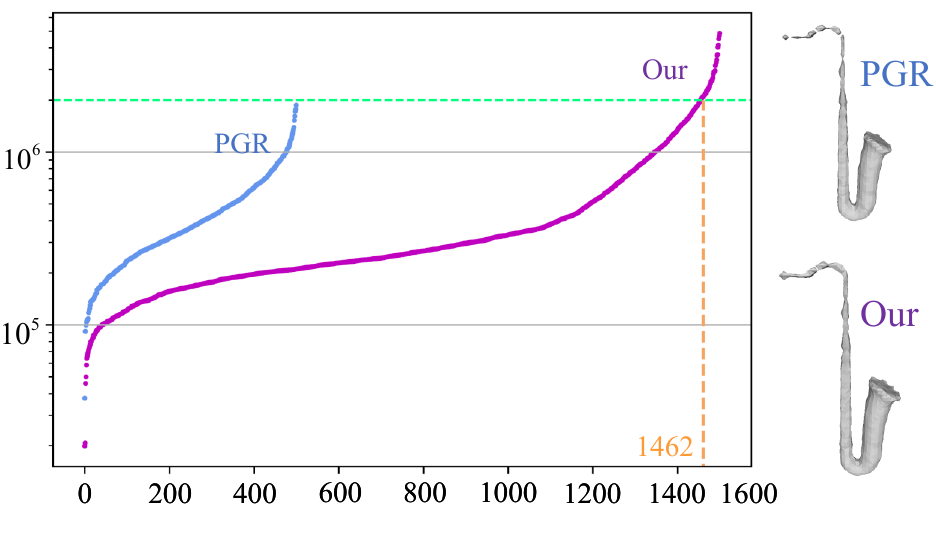}
	\caption{Under the same input point cloud with 500 points and regularization value, our method reduces the singularity of matrix $B$ and increases the number of effective equations compared to PGR.  }
	\label{cond}
\end{figure}

\subsubsection{Solving the Under-determined System}
Storing and solving large-scale linear equation systems remains a challenging problem. The computational complexity can be no less than $O(N\text{log}N)$ even with the fast multipole method (\textbf{FMM}) \cite{DARVE2000195}. Therefore, we tend to solve problems with as few equations as possible. Moreover, as the generalization of PGR, we tend to make our method fully backward compatible with it. 

Then, we introduce the details of solving under-determined equation systems. Take the case of $m=2$, when the coefficient matrix $A$ is $2N\times 3N$ as example,
\begin{equation}\label{4211}
	A \mu=\begin{bmatrix}
		A_{\bm{c}_{1}}(\mathcal{P};\mathcal{P})\\
		A_{\bm{c}_{2}}(\mathcal{P};\mathcal{P})
	\end{bmatrix}\mu=d.
\end{equation}

Therefore, the solution is not unique, and the natural idea is to find the solution with the minimum $L_{2}$ norm,
\begin{align*}
	\mathop{ \min }\limits_{\mu} ||\mu||_{2}, \quad \text{subject to}  \quad A \mu=d.
\end{align*}

Similar to \cite{2022PGR}, it can be computed as
\begin{align}\label{4212}
	\mu = A^{T}\xi ,\quad \text{with} \quad B\xi =d,
\end{align}
where
\begin{align*}
	&B=B_{0}+R_{0},\\
	&B_{0}=AA^{T} ,\quad R_{0}=(\alpha-1)\cdot  \text{diag}(B_{0}),
\end{align*}
and $R_{0}$ acts as a regularization that can be any
symmetric and positive-definite matrix. $\alpha$ is a hyperparameter. Thanks to the increase of anisotropy information in equations, our method can reduce the sensitivity to it, compared to PGR.

As the number of equations increases, calculating $AA^{T}$ directly takes up too much memory. Therefore, we calculate the coefficient matrix $B_{0}$ by blocking matrix, which enables our method to handle point clouds with a larger scale. In detail, we set $N_{s}$ as the batch size. Then $\mathcal{P}=\mathcal{P}_{1}\cup\mathcal{P}_{2}\cup \cdots \cup \mathcal{P}_{M}$ with ${P}_{i}\cap{P}_{j}=\emptyset, \forall i \neq j$. For any $i$,
\begin{equation}\label{4213}
	A_{\bm{c}_{i}}(\mathcal{P};\mathcal{P})=\begin{bmatrix}
		A_{\bm{c}_{i}}(\mathcal{P}_{1};\mathcal{P})\\
		\vdots \\
		A_{\bm{c}_{i}}(\mathcal{P}_{M};\mathcal{P})
	\end{bmatrix}
	\triangleq
	\begin{bmatrix}
		A_{\bm{c}_{i},1}\\
		\vdots \\
		A_{\bm{c}_{i},M}
	\end{bmatrix},
\end{equation}
where $M=[\frac{N}{N_{s}}]+1$. For all $j \leqslant M$, the number of rows in $A_{\bm{c}_{i},j}$ is less than or equal to the hyperparameter $N_{s}$. Then, we use Einstein's summation and the symmetry of $B_{0}$ to calculate and assemble matrix $B_{0}$ in blocks.
\begin{align*}
	(B_{0})_{M*i_{1}+j_{1},M*i_{2}+j_{2}}= A_{\bm{c}_{i_{1}},j_{1}}A_{\bm{c}_{i_{2}},j_{2}}^{T},\quad (M*i_{1}+j_{1} \leqslant M*i_{2}+j_{2}),
\end{align*}
where $i_{1},i_{2}\leqslant m$, $ j_{1},j_{2}\leqslant M$, and $(B_{0})_{i,j}$ are the submatrixes with $(B_{0})_{i,j}=(B_{0})_{j,i}^{T}, \forall i>j$. In this way, we can make the memory required for calculating the coefficient matrix and $B$ linear with the point cloud's size. 

\begin{figure}[htbp]
	\centering
	\includegraphics[width=1.0\linewidth]{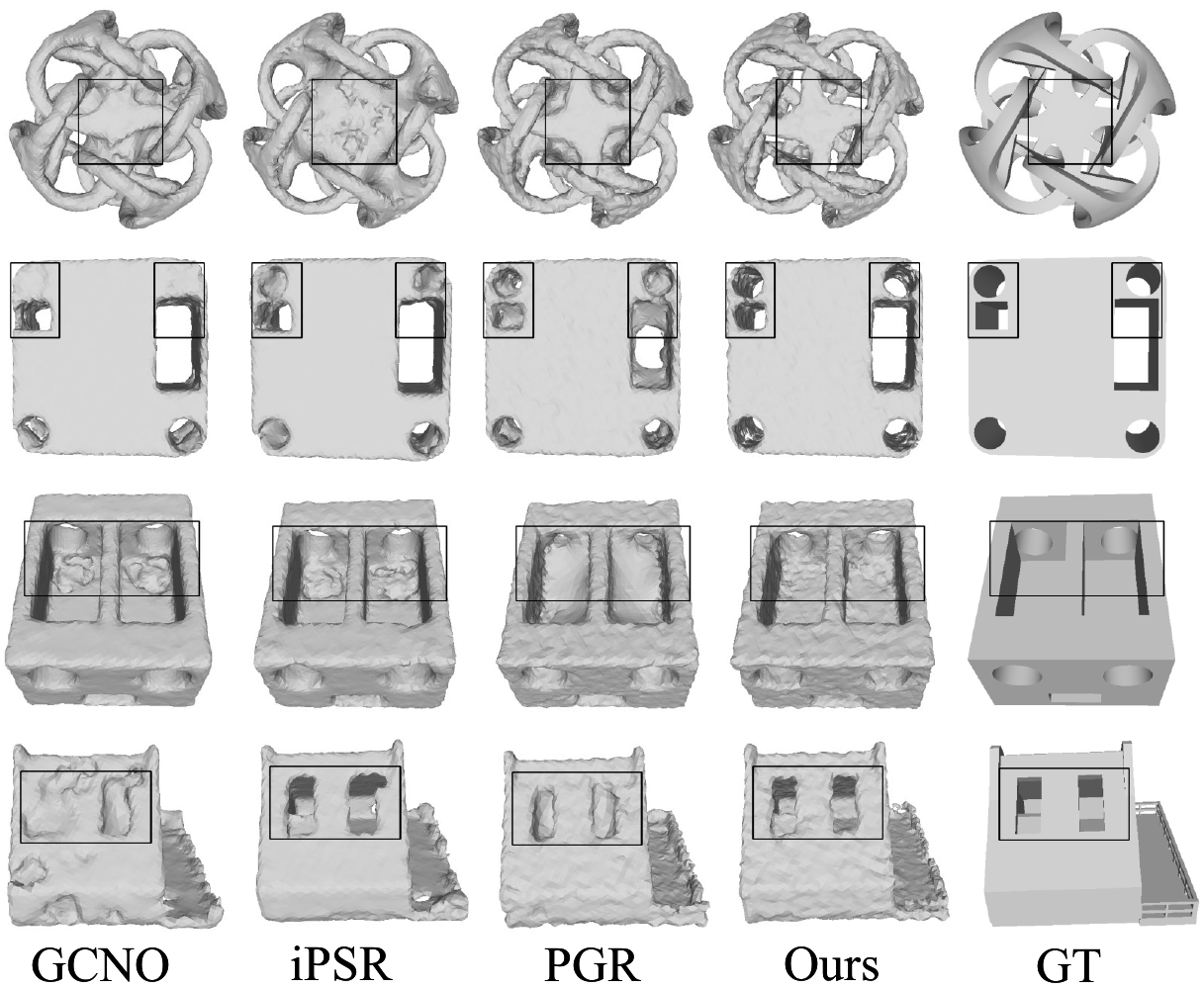}
	\caption{ The qualitative comparison of our method with PGR, iPSR, and GCNO+SPR on reconstruction. Our method can reconstruct surfaces with higher quality.}
	\label{reconstruction}
\end{figure}

\subsubsection{Solving the Over-determined System}
Since the constructed equation system is nonlinear to the velocity vector $\bm{c}$, we can choose infinite velocity vectors with different directional information to construct the equations in theory. In order to fully utilize point cloud information and anisotropy, the idea of letting $m>3$ is natural. At this point, the equations become an over-determined system. We take the case that the coefficient matrix $A$ is $mN\times 3N$ as the example to describe the framework of solving the over-determined system.

At this point, the natural idea is to seek the least-squares solution
\begin{align*}
	\mathop{ \min }\limits_{\mu} ||A\mu-d_{m}||_{2}^{2},
\end{align*}
where $d_{m}=\frac{\bm{1}}{\bm{2}}\in \mathbb{R}^{mN\times 1}$, which has an equivalent form with
\begin{align}\label{4223}
	\mathop{ \min }\limits_{\mu} \sum_{i=1}^{m}||A_{\bm{c}_{i}}\mu-d_{1}||_{2}^{2},
\end{align}
where $d_{1}=\frac{\bm{1}}{\bm{2}}\in \mathbb{R}^{N\times 1}$.
Then, it can be computed as
\begin{align*}
	H_{0}\mu=(\sum_{i=1}^{m}A_{\bm{c}_{i}}^{T}d_{1}), \quad \text{with} \quad H_{0}=(\sum_{i=1}^{m}A_{\bm{c}_{i}}^{T}A_{\bm{c}_{i}}).
\end{align*}

However, $H_{0}$ is also difficult to avoid being ill-conditioned.
For this, we put a similar nonuniform regularization 
\begin{align}\label{4214}
	H\mu=(\sum_{i=1}^{m}A_{\bm{c}_{i}}^{T}d_{1}) \quad \text{with}  \quad H=H_{0}+\text{diag}(H_{0}),
\end{align}

Since $B$ and $H$ are symmetric matrices, we can use the $CG$ algorithm to solve linear equations \eqref{4212} and \eqref{4214}.

Although when $m=3$, the coefficient matrix $A$ still cannot be solved using the $CG$ algorithm directly, either the $GMRES$ algorithm or the methods mentioned above can be used to solve it.

\begin{figure}
	\centering
	\includegraphics[width=1\linewidth]{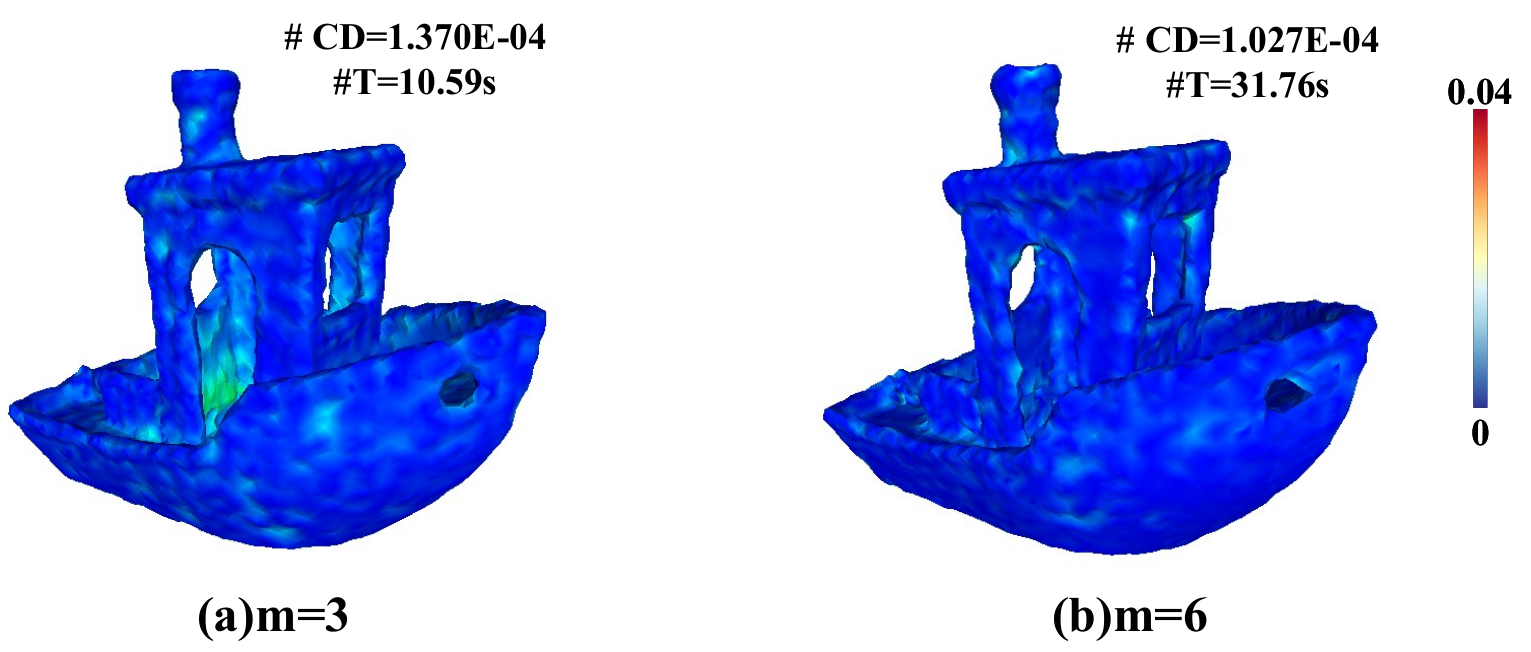}
	\caption{The qualitative and quantitative comparison of equation quantity in our method. Choosing more velocity vectors can further slightly improve the performance.}
	\label{m}
\end{figure}

The qualitative and quantitative results of choosing different numbers $m$ of velocity vectors for reconstruction are compared in Figure \ref{m}. Choosing more velocity vectors can add more anisotropic directional information from the point cloud, increase the number of equations, and improve the performance. Figure \ref{cond} also demonstrates the improvement from the perspective of singular values. For a point cloud with 500 points, we can increase the number of equations to 1462 by choosing three different velocity vectors.

More velocity vectors would significantly increase the computational load while the marginal improvement also decreases. 
Therefore, we fix the number of velocity vectors $m=3$ and solve the equation systems by using \eqref{4212} in experiments.

\subsubsection{Normal Estimation and Orientation}
Orientation and surface reconstruction are closely related in computer graphics. Some proposed state-of-the-art methods (such as iPSR ~\cite{hou2022iterative}, PGR \cite{2022PGR}) can balance orientation and reconstruction. Similarly,  Our method can simultaneously accomplish the orientations and surface reconstruction tasks with state-of-the-art performance.
As shown in \eqref{LSE}, our method can output the points' unit outward normals by normalizing the \textbf{LSE} because the area element only determines the module length instead of the direction.

Firstly, we reshape $\mu$ into the matrix with the size of $N\times 3$, where each row represents an input point's normal information,
\begin{align*}
	\mu\in \mathbb{R}^{3N\times 1} \to \widehat{\mu}=\begin{bmatrix}
		\mu_{1}&\mu_{2}&\mu_{3}\\
		\vdots&\vdots&\vdots\\
		\mu_{3N-2}&\mu_{3N-1}&\mu_{3N}\\
	\end{bmatrix}\in \mathbb{R}^{N\times 3}.
\end{align*}

Then, we can complete the normal estimation and orientation of the point cloud by normalizing vectors row by row. Let $\widehat{\bm{n}}_{\bm{p}_{i}}$ be the estimated normal for point $\bm{p}_{i} \in \mathcal{P}$,
\begin{align*}
	\widehat{\bm{n}}_{\bm{p}_{i}}=\dfrac{(\mu_{3i+1},\mu_{3i+2},\mu_{3i+3})}{\sqrt{\mu_{3i+1}^{2}+\mu_{3i+2}^{2}+\mu_{3i+3}^{2}}}.
\end{align*}

Figures \ref{orientation1} and  \ref{orientation2} show the qualitative comparison of normal estimation and orientation. Our method stimulates the potential for orientation in the Gauss formula by introducing anisotropy, especially for point clouds with thin structures or small holes. Our method achieves a high-quality, consistent orientation and the smaller deviation angle between the estimated normals and the ground truth, exhibiting superior results compared to other state-of-the-art methods.

\subsubsection{Iso-surface Extraction}
After obtaining $\mu$, we compute the value of the indicator function on the 
query point set $\mathcal{Q}=\{\bm{q}_{s}\}_{s=1}^{N_{Q}}$, which is the set of all corner points on the octree, by the average of the matrices multiplication
\begin{equation*}
	\dfrac{1}{m}\sum_{i=1}^{m}A_{\bm{c}_{i}}(\mathcal{Q};\mathcal{P})\mu.
\end{equation*}

We follow previous works (i.e. \cite{kazhdan2013screened}, \cite{2019GR}) to extract surfaces with marching cubes on octrees. Note that the regularization term $\alpha$  in equation \eqref{4212} or \eqref{4214} may cause the iso-value's slight shift from $\frac{1}{2}$. We update the iso-value $v_{\text{iso}}$ with 
\begin{align*}
	v_{\text{iso}}&=\text{average}(\dfrac{1}{m}\sum_{i=1}^{m}A_{\bm{c}_{i}}(\mathcal{P};\mathcal{P})\mu)\\
	&=\dfrac{1}{mN}\sum_{i=1}^{m}\sum_{j=1}^{N}A_{\bm{c}_{i}}(\bm{p}_{j};\mathcal{P})\mu.
\end{align*}

Our method can demonstrate the state-of-the-art performance of reconstruction from Figures \ref{reconstruction}, \ref{thin1}, and  \ref{holes}, which show the qualitative comparison of reconstructions with other well-known methods.

\begin{algorithm}[htbp]
	\SetAlgoLined
	\caption{Anisotropy-based Gauss Reconstruction}\label{algorithm2}
	\KwData{Unoriented point cloud $\mathcal{P}$, maximum depth of octree $D_{\max}$, width function
		parameters $w_{\min}$,$w_{\max}$, regularization scaling factor $\alpha$, batch size $N_{s}$, the modulus of velocity vectors $L$, the group of input velocity vectors $\bm{C}=(\bm{c}_{1},\bm{c}_{2},\cdots,\bm{c}_{m})$.}
	\KwResult{Oriented point cloud $(\mathcal{P},\mathcal{N})$, a watertight surface $M$ approximating the points.}
	Set up octree $O$ of max depth $D_{\max}$ by $\mathcal{P}$ and
	obtain the corner points of the octree as query points $\mathcal{Q}$ \;
	
	Compute the width of the input points as ~\citeN{2019GR} \;
	
	Calculate the eigenvalues $\lambda_{1}$, $\lambda_{2}$, $\lambda_{3}$ and corresponding eigenvectors of point clouds $\bm{v}_{1}$, $\bm{v}_{2}$, $\bm{v}_{3}$\;
	
	Based on whether $\lambda_{3}\leqslant 0.001$ to calculate velocity vectors $\bm{c}_{a_1},\bm{c}_{a_2},\bm{c}_{a_3}$ by adaptive strategy and decide whether to add to $\bm{C}$. If added, $\bm{C}\leftarrow \bm{C}\cup \bm{c}_{a_1}\cup\bm{c}_{a_2}\cup\bm{c}_{a_3}$, $m\leftarrow m+3$\;
	
	\If{$m\leqslant 3$}
	{
		\For{$i = 1$ \KwTo $m$}
		{
			Calculate matrix $A_{\bm{c}_{i}}$ by dividing rows into blocks based on batch size $N_{s}$, i.e
			$A_{\bm{c}_{i}}$=$(A_{\bm{c}_{i,1}},A_{\bm{c}_{i,2}},\cdots,A_{\bm{c}_{i,s}})^{T}$\;
			\For{$j = 1$ \KwTo $m$}
			{
				Calculate $A_{\bm{c}_{j}}$  by dividing rows into blocks similarly\;
				Using Einstein summation to calculate $B_{i,j}$\;
			}
		}
		Concatenate the complete matrix $B$ and apply regularization $B \leftarrow B+(\alpha-1)\cdot \text{diag} (B)$ \;
		
		Solve $B\xi=\bm{\frac{1}{2}}$ for $\xi$ by conjugate gradient(CG) algorithm \;
		
		Calculate the linearized surface element $\mu=A^{T} \xi$ by Equation \eqref{4212} \;
	}
	
	\If{$m\geqslant 3$}
	{
		\For{$i = 1$ \KwTo $m$}
		{
			Calculate matrix $A_{\bm{c}_{i}}$ and $B \leftarrow B+A_{\bm{c}_{i}}^{T}A_{\bm{c}_{i}}$, $d \leftarrow d+A_{\bm{c}_{i}}^{T} \bm{\frac{1}{2} }$\;
		}
		Concatenate the complete matrix $B$ and apply regularization $B \leftarrow B+(\alpha-1)\cdot \text{diag} (B)$ \;
		Calculate the linearized surface element $\mu$ with $B\mu=d$ by conjugate gradient(CG) algorithm \;
	}
	
	Extract normals from $\mu$ and output oriented point cloud $(\mathcal{P},\mathcal{N})$ 
	Calculate  the average coefficient matrix $\overline{A}=(\sum_{k=1}^{m}A_{\bm{c}_{k}})/m$ \;
	Compute the iso-value as $v_{\text{iso}} = \text{average}(\overline{A}(\mathcal{P}; \mathcal{P})\mu)$ \;
	Obtain query points’ values as $\overline{A}(\mathcal{Q}; \mathcal{P})\mu$ \;
	Reconstruct surface $M$ using marching cubes by
	$\overline{A}(\mathcal{Q}; \mathcal{P})\mu$ and $v_{\text{iso}}$ \;
\end{algorithm} 

\subsection{Adaptive Selection of Velocity Vectors}
We select velocity vectors based on the singular value decomposition.



First, we select a subset $\mathcal{P}_{1}$ of the input point cloud $\mathcal{P}$ for acceleration, calculate the geometric center $\bm{c}=(c_{x},c_{y},c_{z})$ by
\begin{align*}
	\bm{c}= \dfrac{1}{|\mathcal{P}_{1}|}\sum_{i=1}^{|\mathcal{P}_{1}|}\bm{p}_{i},
\end{align*}
and centralized points to $\bm{p}'_{i}=(p'_{i,x},p'_{i,y},p'_{i,z}),i=1,2,\cdots |\mathcal{P}_{1}|$,  with
\begin{align*}
	\bm{p}'_{i}=\bm{p}_{i}- \bm{c}.
\end{align*}

Secondly, we can establish the corresponding covariance matrix
\begin{align*}
	Cov=\dfrac{1}{|\mathcal{P}_{1}|}\sum_{i=1}^{|\mathcal{P}_{1}|}
	\begin{bmatrix}
		p'_{i,x} \cdot p'_{i,x} & p'_{i,x} \cdot p'_{i,y} & p'_{i,x} \cdot p'_{i,z}\\
		p'_{i,y} \cdot p'_{i,x} & p'_{i,y} \cdot p'_{i,y} & p'_{i,y} \cdot p'_{i,z}\\
		p'_{i,z} \cdot p'_{i,x} & p'_{i,z} \cdot p'_{i,y} & p'_{i,z} \cdot p'_{i,z}\\
	\end{bmatrix},
\end{align*}
where $|\cdot|$ represents the number of elements in the set, and  $Cov \in \mathbb{R}^{3 \times 3}$ is the symmetric matrix. Then, we can perform eigenvalue decomposition on the covariance matrix by spectral theorem
\begin{align*}
	Cov=V\begin{bmatrix}
		\lambda_{1}&&\\
		&\lambda_{2}&\\
		&&\lambda_{3}\\
	\end{bmatrix}V^{T},
\end{align*}
where $\lambda_{1}\geqslant \lambda_{2} \geqslant \lambda_{3}$. $V=(\bm{v}_{1},\bm{v}_{2},\bm{v}_{3})^{T}$ is the matrix composed of the respective orthogonal eigenvectors, and the eigenvalues $\lambda_{i}$ represent the degree of concentration of point cloud distribution in the corresponding direction $\bm{v}_{i}$.

\begin{figure*}
	\centering
	\includegraphics[width=1.0\linewidth]{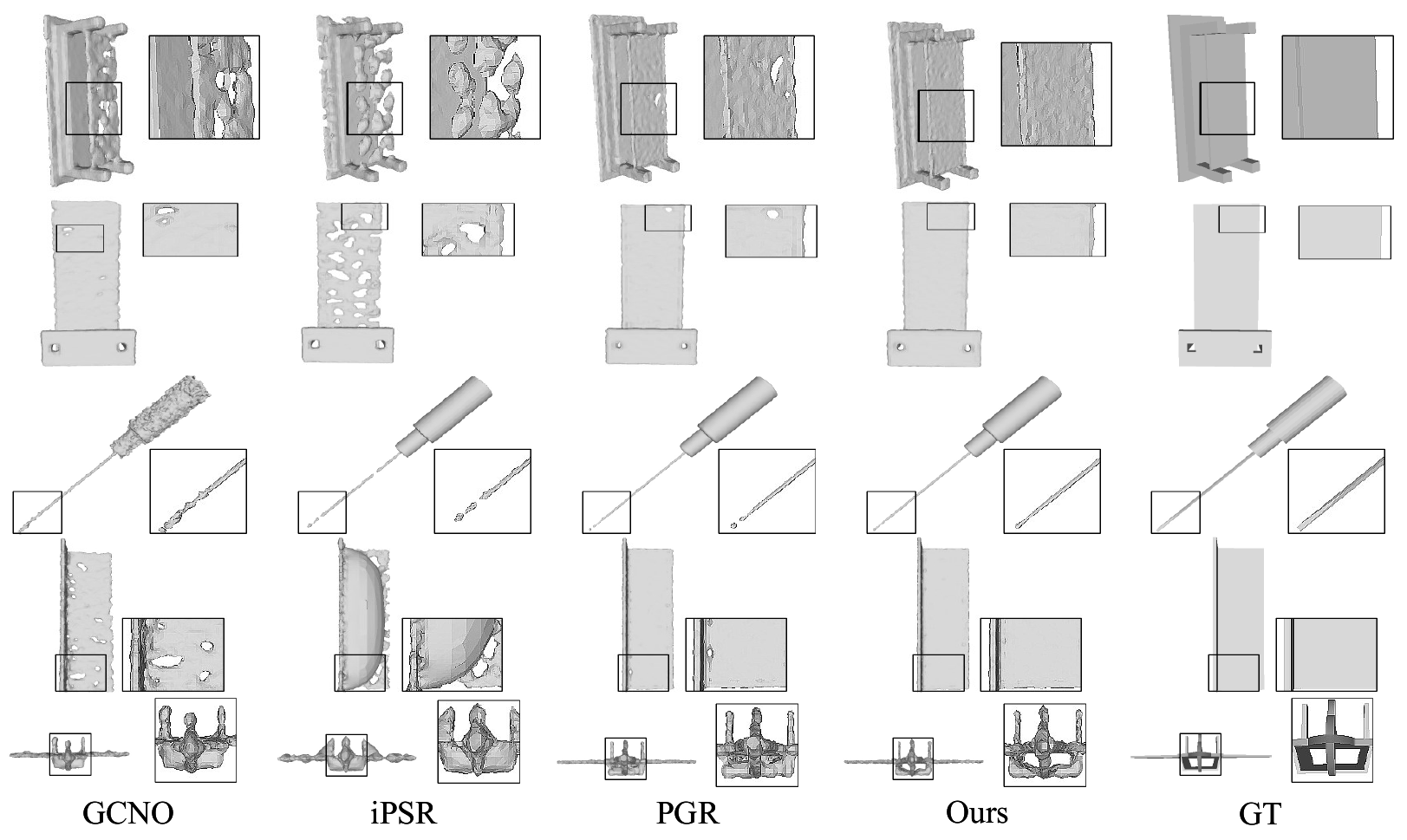}
	\caption{Qualitative comparison of our method with other state-of-the-art methods on the reconstructions from point clouds with thin structure(the first, second, and fourth row), needle tip (the third row), or thin-walled space (the fifth row). Other state-of-the-art methods often fall into the issues, including sheet breakage, tip discontinuities, and wrong sealing. However, our method can handle these issues.}
	\label{thin1}
\end{figure*}

Hence, $\bm{v}_{3}$ corresponds to the minimum eigenvalue's eigenvector and represents the plate's normal. In addition, we can determine the structural characteristics of point clouds by observing the eigenvalues $\lambda_{3}$ by the fact that the proximity of $\lambda_{3}$ to zero indicates a higher degree of thin structure within the corresponding point cloud.

Finally, based on the above analysis, we can complete the identification by whether it meets $\lambda_{3} \leqslant \varepsilon$, where $\varepsilon$ is a hyperparameter. Since the point cloud is standardized into the bounding box $[0,1]^{3}$, $\varepsilon$ is set to the constant $0.001$ in the experiments in this article.


Then, the velocity field is chosen as follows.
\begin{equation*}
	\left \{\begin{aligned}
		&\bm{c}_{1}=L \cdot \bm{v}_{1}, \bm{c}_{2}=L \cdot \bm{v}_{2}, \bm{c}_{3}=L \cdot \bm{v}_{3},& &  \lambda_{3}  >\varepsilon\\
		&\bm{c}_{1}=L \cdot \bm{v}_{1}, \bm{c}_{2}=L \cdot \bm{v}_{2},\bm{c}_{3}=\frac{2\varepsilon L}{\lambda_{3}+0.1\varepsilon} \cdot \bm{v}_{3} ,& &    \lambda_{3} \leqslant \varepsilon,
	\end{aligned} \right.
\end{equation*}
where $L$ is a given hyperparameter. The default value of $L$ is 1.0, and the recommended setting range is 0.5 to 6. Algorithm \ref{algorithm2} gives the detailed description of our method.

\begin{figure}[htbp]
	\centering
	\includegraphics[width=1.0\linewidth]{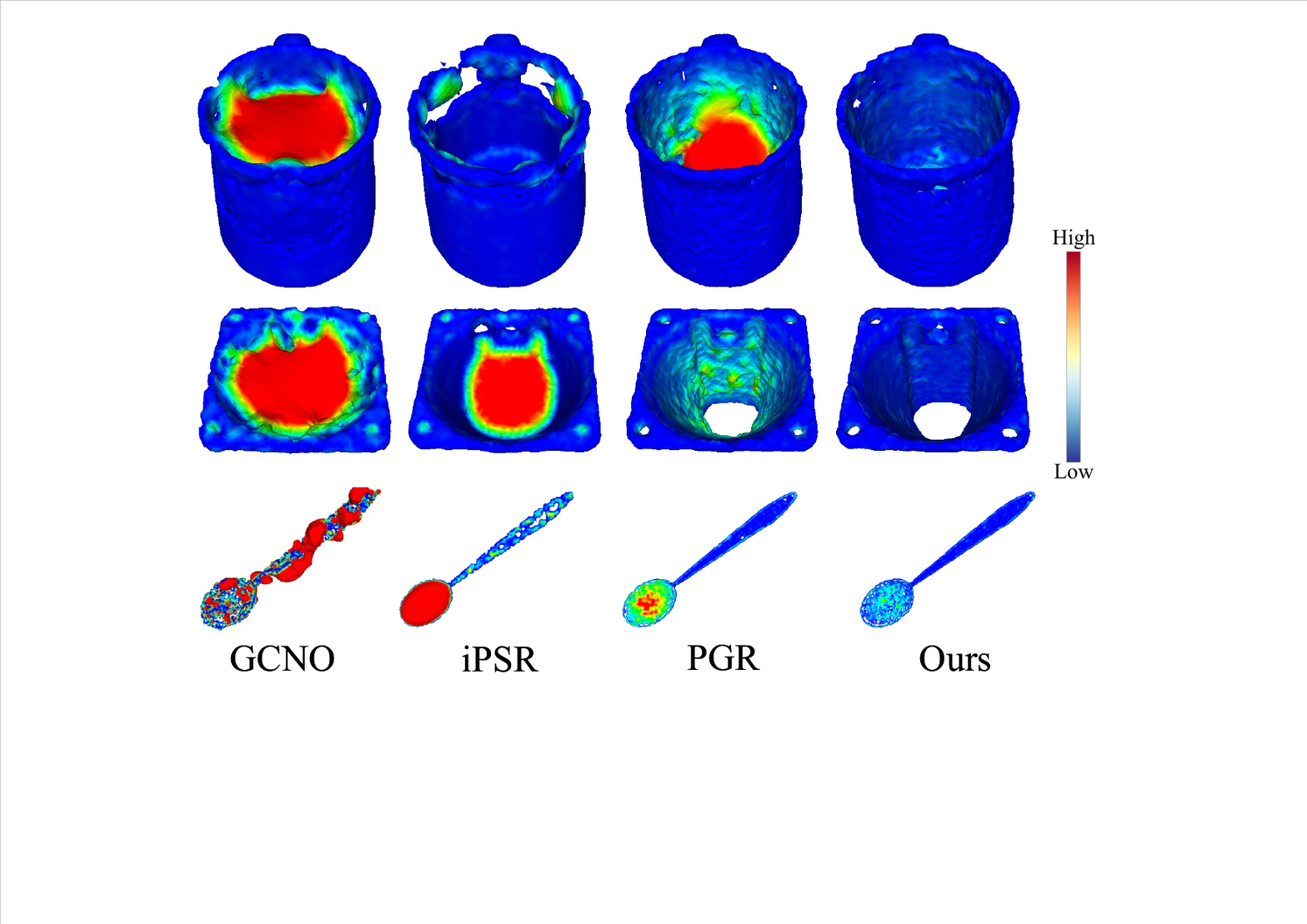}
	\caption{ Error colormaps of the cup in the Famous dataset, triangular pyramid in the Thingi10K dataset, and spoon in the Thin dataset. The colors in the figure represent the degree of error. Compared to the wrong sealing of other well-known methods, our method demonstrates state-of-the-art reconstruction performance.}
	\label{21}
\end{figure}

\begin{figure}[htbp]
	\centering
	\includegraphics[width=1.0\linewidth]{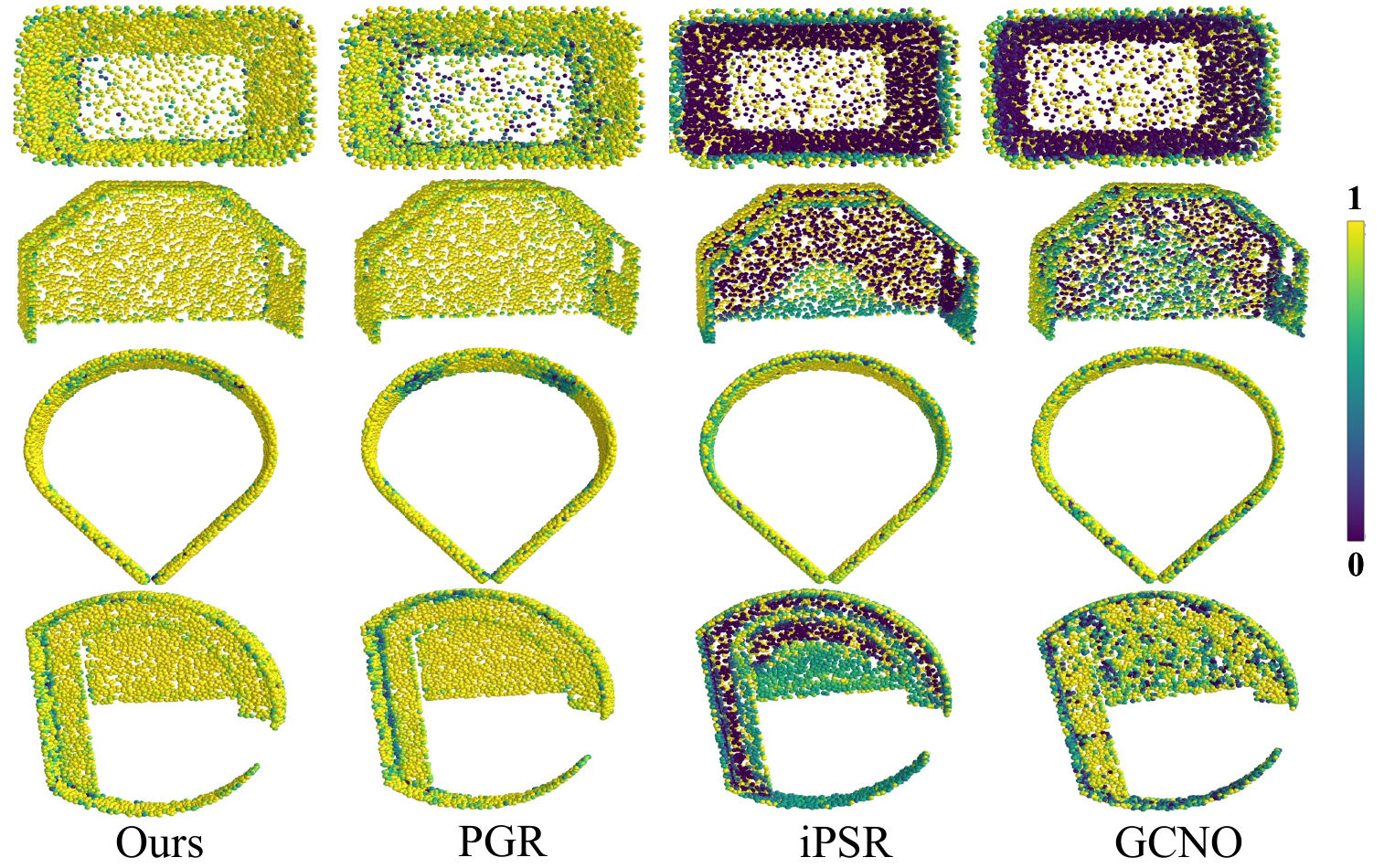}
	\caption{Qualitative comparison of our method with other state-of-the-art methods about normal estimation. The color of the points represents the inner product of the estimated normals and the ground truth.}
	\label{orientation1}
\end{figure}

For point clouds with strong anisotropy, our method outperforms results as shown in Figures \ref{21},\ref{thin1},\ref{orientation1}, and the supplementary materials. 

\section{Experiments}\label{sec:6}
\begin{table*}[tbph]\tiny
	\centering
	
	\begin{tabular}{|l|ccccc|ccccc|ccccc|ccccc|}
		\hline
		\multirow{2}{*}{Models}& \multicolumn{5}{c|}{$\text{PGP}_{90}$  $\uparrow $} & \multicolumn{5}{c|}{$N_{p}$ $\uparrow$}& \multicolumn{5}{c|}{$N_{s}$ $\uparrow$}& \multicolumn{5}{c|}{ CD  $\downarrow$ } \\
		
		& Ours  & PGR & iPSR & GCNO & PCA & Ours  & PGR & iPSR & GCNO & PCA & Ours  & PGR & iPSR & GCNO & PCA & Ours  & PGR & iPSR & GCNO & PCA \\
		\hline

		ABC00019908 & \textbf{0.9940} & 0.9930 & 0.9901 & 0.9736 & 0.8476 & \textbf{0.9288} & 0.9258 & 0.9265 & 0.8638 & 0.7486 & \textbf{0.9687} & 0.9662 & 0.9664 & 0.8376 & 0.7365 & \textbf{3.13} & 3.31  & 3.16  & 8.13  & 13.48 \\
		ABC00015485 & \textbf{1.0000} & \textbf{1.0000} & \textbf{1.0000} & 0.9930 & 0.9058 & \textbf{0.9621} & 0.9595 & 0.9527 & 0.9381 & 0.7911 & \textbf{0.9756} & 0.9740 & 0.9744 & 0.9238 & 0.7902 & \textbf{6.54} & 8.25  & 6.71  & 8.95  & 18.02 \\
		Trash\_can20 & \textbf{0.9690} & 0.4975 & 0.5385 & 0.5200 & 0.5004 & \textbf{0.8378} & 0.1373 & 0.0722 & 0.0218 & 0.0036 & \textbf{0.7868} & 0.3179 & 0.0820 & 0.0809 & 0.0285 & \textbf{142.98} & 235.99 & 304.09 & 275.71 & 2654.13 \\
		Spoon5 & \textbf{1.0000} & \textbf{1.0000} & 0.9995 & 0.5336 & 0.5122 & \textbf{0.9563} & 0.9493 & 0.6066 & 0.0496 & 0.0181 & \textbf{0.9042} & 0.8169 & 0.7308 & 0.2336 & 0.0178 & \textbf{0.79} & 1.03  & 1.75  & 118.78 & 2996.84 \\
		Thingi10k10218 & \textbf{0.9923} & 0.9919 & 0.9816 & 0.9434 & 0.8752 & \textbf{0.9330} & 0.9318 & 0.9234 & 0.8378 & 0.7205 & \textbf{0.9652} & 0.9553 & 0.9618 & 0.7964 & 0.8809 & \textbf{9.09} & 12.16 & 9.12  & 35.03 & 43.95 \\
		Screwdriver16 & \textbf{0.9804} & 0.9802 & 0.9793 & 0.7033 & 0.9504 & \textbf{0.9260} & 0.9250 & 0.9144 & 0.3627 & 0.8846 & \textbf{0.9563} & 0.9547 & 0.9510 & 0.6685 & 0.8331 & \textbf{1.82} & 1.97  & 1.89  & 2.76  & 53.74 \\
		Plate & \textbf{1.0000} & \textbf{1.0000} & \textbf{1.0000} & 0.9998 & 0.5594 & 0.9575 & 0.9458 & 0.8659 & \textbf{0.9628} & 0.1321 & 0.9722 & 0.9644 & 0.9335 & \textbf{0.9750} & 0.0436 & \textbf{5.95} & 6.62  & 7.12  & 6.14  & 341.23 \\
		Thingi10k16680 & \textbf{0.9966} & 0.9932 & 0.9965 & 0.6424 & 0.5093 & \textbf{0.9339} & 0.9095 & 0.9326 & 0.2229 & 0.0036 & \textbf{0.9661} & 0.9100 & 0.9639 & 0.5697 & 0.0586 & \textbf{15.69} & 54.78 & 16.11 & 282.61 & 1256.29 \\
		Thingi10k88053 & \textbf{1.0000} & \textbf{1.0000} & 0.5595 & 0.9633 & 0.9891 & \textbf{0.9687} & 0.9597 & 0.1208 & 0.8927 & 0.9679 & \textbf{0.9815} & 0.8914 & 0.2227 & 0.7959 & 0.9548 & \textbf{3.88} & 10.89 & 19.38 & 7.68  & 49.38 \\
		Cup34 & \textbf{1.0000} & 0.9945 & 0.9995 & 0.8156 & 0.6636 & \textbf{0.9363} & 0.9282 & 0.9256 & 0.6147 & 0.3266 & \textbf{0.9830} & 0.7872 & 0.9586 & 0.6096 & 0.4307 & \textbf{13.35} & 300.74 & 16.44 & 142.59 & 254.35 \\
		Utah\_teapot & \textbf{0.9950} & 0.9945 & 0.9869 & 0.9920 & 0.9878 & \textbf{0.9499} & 0.9483 & 0.9438 & 0.9284 & 0.9439 & \textbf{0.9742} & 0.9711 & 0.9734 & 0.8895 & 0.9732 & \textbf{4.23} & 4.49  & 4.36  & 6.97  & 4.50 \\
		Thingi10k132420 & \textbf{1.0000} & \textbf{1.0000} & \textbf{1.0000} & \textbf{1.0000} & \textbf{1.0000} & \textbf{0.9879} & 0.9852 & 0.9825 & 0.9679 & 0.9691 & \textbf{0.9900} & 0.9898 & 0.9882 & 0.9727 & 0.9702 & \textbf{0.96} & 0.97  & 0.98  & 0.97  & 1.61 \\
		Saxophone2 & \textbf{1.0000} & \textbf{1.0000} & \textbf{1.0000} & 0.6047 & 0.9332 & \textbf{0.9709} & 0.9702 & 0.9706 & 0.1359 & 0.8620 & \textbf{0.9840} & 0.9736 & 0.9813 & 0.1039 & 0.8099 & \textbf{1.12} & 1.18  & 1.46  & 4873.62 & 161.75 \\
		\hline
		Average & \textbf{0.9944} & 0.9573 & 0.9255 & 0.8219 & 0.7872 & \textbf{0.9422} & 0.8827 & 0.7798 & 0.5999 & 0.5670 & \textbf{0.9544} & 0.8825 & 0.8221 & 0.6505 & 0.5791 & \textbf{16.12} & 49.41 & 30.18 & 443.84 & 603.79 \\
		\hline
	\end{tabular}%
	\caption{Comparison of our method with other state-of-the-art methods for point clouds with 10K points on orientation and reconstruction. The CD values are multiplied by $10^{5}$. Our method exhibits superior performance against other state-of-the-art methods.}
	\label{10K}
\end{table*}%

\paragraph{\textbf{Experimental Setup}}
Experiments are conducted using an NVIDIA GeForce RTX 3090 graphics card with 24GB video memory. The process of solving equations involves large matrix multiplications, for which we use Cupy \cite{nishino2017cupy} for high-efficiency matrix computation on the GPU, like PGR. 

\paragraph{\textbf{Evaluating Indicator}}
The proportion of good points (PGP)  is the ratio of points for which the angle between the estimated normal and its corresponding ground truth normal is smaller than a specified threshold. The higher value of PGP indicates better consistency in the orientation of the point cloud.
\begin{equation*}
	\text{PGP}_{90}(P)=|\text{correct}.P|/|P|, \text{correct}.P=\{\bm{p}_{i} \in P | \  \bm{n}_{\bm{p}_{i,\text{out}}} \cdot \bm{n}_{\bm{p}_{i,\text{true}}}>0\}.
\end{equation*}

The Chamfer distance (CD)  penalizes both false negatives (missing parts) and false positives (excess parts) to evaluate the reconstruction error.
\begin{equation*}
	\text{CD}(S_{1},S_{2})=\dfrac{1}{|S_{1}|}\sum_{\bm{x} \in S_{1}}\mathop{\min }\limits_{\bm{y} \in S_{2}}||\bm{x}-\bm{y}||_{2}^{2}+\dfrac{1}{|S_{2}|}\sum_{\bm{y} \in S_{2}}\mathop{\min }\limits_{\bm{x} \in S_{1}}||\bm{x}-\bm{y}||_{2}^{2}.
\end{equation*}

$S_{1}$ and $S_{2}$ denote the reconstructed and ground truth surfaces respectively. In the set in the subsection of evaluating indicators, $|\cdot|$ represents the number of elements. To evaluate the quality of the reconstruction, we employ a sampling approach using 20,000 points for both surfaces. 

Normal consistency  (expressed as a percentage and abbreviated as ‘NC’) reflects the degree of normal consistency between two point clouds. The NC value is computed as follows:
\begin{align}\label{156}
	\small
	&\text{NC}(P_{1}, P_{2}) =\frac{1}{2|P_{1}|}\sum_{\bm{p}_{1} \in P_{1}}\bm{n}_{\bm{p}_{1}}\bm{n}_{\text{closest}(\bm{p}_{1},P_{2})}+\frac{1}{2|P_{2}|}\sum_{\bm{p}_{2} \in P_{2}} \bm{n}_{\bm{p}_{2}}\bm{n}_{\text{closest}(\bm{p}_{2},P_{1})} \nonumber\\ 
	&\text{closest}(\bm{p},P)=\mathop{\arg \min }\limits_{\bm{p}' \in P} d(\bm{p},\bm{p}').
\end{align}

$N_{p}$ means that $P_{1}$ is chosen as the ground truth of the input point cloud, and $P_{2}$ is the output oriented point cloud by the algorithm. As a supplement to $\text{PGP}_{90}$, it is mainly used to measure the algorithm's ability of orientation and normal estimation.

$N_{s}$ means that $P_ {1}$ is sampled from the surface of the ground truth, $P_ {2}$ is sampled from the output surface by the algorithm as a supplement to the Chamfer distance, it is mainly used to measure the algorithm's ability of surface reconstruction.

\begin{table}
	\centering
	\caption{Quantitative comparisons of our method with other state-of-the-art methods on the orientation and reconstruction in the clean datasets. The CD values are multiplied by $10^{5}$. Our method demonstrates state-of-the-art performance.}
	\label{table2}
	
	\centering
	\begin{tabular}{cccccc}
		\toprule[1.2pt]
		\multicolumn{2}{c}{} & Real-world  & Famous & ABC   & Thingi10k \\
		\midrule[1.2pt]
		\multirow{4}{*}{$\text{PGP}_{90}$ $\uparrow $} & GCNO  &   0.9537    &   0.9454    &  -     & - \\
		& iPSR  & 0.9747 & 0.9772 & 0.9128 & 0.9685 \\
		& PGR   & 0.9894 & 0.9776 & 0.9592 & 0.9859 \\
		& Ours  & \textbf{0.9914} & \textbf{0.9802} & \textbf{0.9661} & \textbf{0.9879} \\
		\hline
		\multirow{4}{*}{\centering $\text{NC}_{p}$ $\uparrow $} & GCNO  &   0.7842    &   0.8196    &     -    &   - \\
		& iPSR  & 0.8488 & 0.8447 & 0.7614 & 0.8646 \\
		& PGR   & 0.8893 & 0.8440 & 0.8370 & 0.8986 \\
		& Ours  &\textbf{0.8899} & \textbf{0.8487} & \textbf{0.8458} & \textbf{0.9011} \\
		\hline
		\multirow{4}{*}{$\text{NC}_{s}$ $\uparrow $} & GCNO  &   0.8244    &  0.8447     &     -    &  -  \\
		& iPSR  & 0.8852 & 0.8970 & 0.7883 & 0.8983 \\
		& PGR   & 0.9038 & 0.8779 & 0.8319 & 0.9122 \\
		& Ours  & \textbf{0.9193} & \textbf{0.8991} & \textbf{0.8660} & \textbf{0.9277} \\
		\hline
		\multirow{4}{*}{CD $\downarrow $} & GCNO  &   41.38     &  36.83    &     -    &  -  \\
		& iPSR   & 28.02 & 21.77 & 45.18 & 18.10\\
		& PGR   & 30.43 & 9.46  & 40.40  & 18.24 \\
		& Ours  & \textbf{25.90} & \textbf{8.30} & \textbf{15.86} & \textbf{12.23} \\
		\bottomrule[1.2pt]
	\end{tabular}%
	\label{table_MAP}
\end{table}

\begin{table}
	\centering
	\caption{Quantitative comparisons of our method with other state-of-the-art methods on the orientation and reconstruction in the datasets with  0.5\% Gaussian noise. The CD values are multiplied by $10^{5}$. Our method exhibits superior performance against other state-of-the-art methods in orientation and reconstruction.}
	\label{T3}
	\centering
	\begin{tabular}{cccccc}
		\toprule[1.2pt]
		\multicolumn{2}{c}{} & Real-world  & Famous & ABC   & Thingi10k \\
		\midrule[1.2pt]
		\multirow{4}{*}{$\text{PGP}_{90}$ $\uparrow $} & GCNO  &   0.8864    &   0.9001    &   -     &  -  \\
		& iPSR  &0.9411 & 0.9097 & 0.8487 & 0.9509 \\
		& PGR  & 0.9412 & 0.9346 & 0.9180 & 0.9487  \\
		& Ours  &\textbf{0.9427} & \textbf{0.9393} & \textbf{0.9216} & \textbf{0.9615} \\
		\hline
		\multirow{4}{*}{\centering $\text{NC}_{p}$ $\uparrow $} & GCNO  &   0.7201    &  0.6875     &    -     &  -  \\
		& iPSR  & 0.7566 & 0.6908 & 0.6027 & 0.8108  \\
		& PGR   & 0.7343 & 0.7203 & 0.7109 & 0.8084  \\
		& Ours  & \textbf{0.7592} & \textbf{0.7241} & \textbf{0.7147} & \textbf{0.8168} \\
		\hline
		\multirow{4}{*}{$\text{NC}_{s}$ $\uparrow $} & GCNO  &  0.7795     &   0.8335    &    -     &  -  \\
		& iPSR  & 0.8587 & 0.8674 & 0.7586 & 0.8765  \\
		& PGR   & 0.9011 & 0.8773 & 0.7964 & 0.8766 \\
		& Ours  & \textbf{0.9137} & \textbf{0.8860} & \textbf{0.8347} & \textbf{0.8834}  \\
		\hline
		\multirow{4}{*}{CD $\downarrow $} & GCNO  &  61.38     &   39.68    &     -    &  -  \\
		& iPSR   & 73.38 & 22.96 & 46.65 & 19.64 \\
		& PGR   & 31.89 & 9.68 & 46.21 & 18.91 \\
		& Ours  & \textbf{26.28} & \textbf{8.46} & \textbf{19.18} & \textbf{12.71}   \\
		\bottomrule[1.2pt]
	\end{tabular}%
	
\end{table}

\paragraph{\textbf{Parameters}}
We adopt the parameter setting: $w_{\text{min}}=0.0015$, $L=1.0$, $N_{s}=5000$ in the experiments and use the CG
algorithm implemented in Python for solving the equation systems. We set the regularization term $\alpha$ to $1.2-2.0$ for the clean point clouds and $2.0 - 5.0$ for the noisy point clouds.

\paragraph{\textbf{Baselines}} We include three well-known and state-of-the-art (SOTA) methods (PGR \cite{2022PGR}, GCNO \cite{2023GCNO}, iPSR \cite{hou2022iterative}) for comparison.  For
GCNO \cite{2023GCNO}, we follow the default setting and match it with SPR \cite{kazhdan2013screened} for reconstruction. Due to the slow running speed of GCNO, we do not test its performance on large datasets. For PGR \cite{2022PGR} and iPSR \cite{hou2022iterative}, besides experimenting with default parameters, we attempt to find the optimal parameters and display the optimal results. In most qualitative examples, our method adopts the same parameter values as PGR.

\subsection{Experimental Effect}
\paragraph{\textbf{Famous, ABC and Thingi10K datasets}}
The Famous ~\citep{erler2020points2surf} dataset includes dozens of classic shapes such as bunny, dragon, and armadillo. The ABC ~\citep{kingma2014adam} dataset comprises a diverse collection of CAD meshes, while the Thingi10K ~\citep{zhou2022learning} dataset contains a variety of shapes with intricate geometric details. We randomly
sample 5K points from each mesh. The quantitative comparison results of the methods and other well-known methods are shown in Table \ref{table2}.
We further tested our method's ability and performance for dense point clouds that reflect the method's ability to reconstruct the details of the structure. Table \ref{10K} shows the results compared with other well-known methods. The comparison of the qualitative results of normal estimation and orientation on these datasets is shown in Figure \ref{orientation1}, while the comparison of qualitative results of surface reconstruction on these datasets is shown in Figures \ref{reconstruction} and \ref{21}.
Experimental results demonstrate the state-of-the-art performance of our method in these datasets.

\paragraph{\textbf{Real-world Dataset}}
The Real-world ~\citep{hou2022iterative} dataset contains noise and outliers from the scan. In addition, the ground truth exhibits lower smoothness. To verify
the algorithm’s robustness, we utilize the Real-world dataset to generate point clouds with 5K points as input. The quantitative comparison results of the methods are shown in Table \ref{table2}.

\begin{figure}[thbp]
	\centering
	\includegraphics[width=1\linewidth]{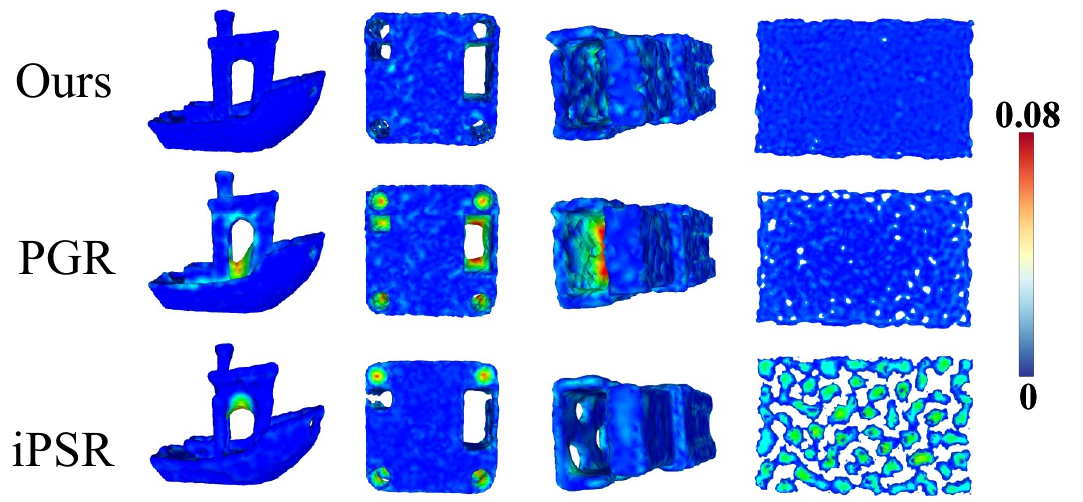}
	\caption{Qualitative results of the reconstruction of noisy point clouds. Especially compared to PGR and iPSR, our method can better protect holes and avoid sealing them.}
	\label{noisy}
\end{figure}

\paragraph{\textbf{Noisy Datasets}}
The performance on noisy datasets can provide compelling evidence of the ability to handle variations and disturbances. However, most non-learning methods cannot handle  
noisy point clouds well, especially sparse point clouds. To evaluate this capability of our method, we introduce a uniform Gaussian noise of 0.5\% to the Famous ~\citep{erler2020points2surf}, ABC ~\citep{kingma2014adam}, Thingi10K ~\citep{zhou2022learning} and Real-world ~\citep{hou2022iterative} datasets with 5K points as input. The quantitative comparison results of our method and other well-known methods are shown in Table \ref{T3}. Figure \ref{noisy} displays the reconstructions from noisy point clouds. Numerous experiments showcase our method's strong robustness and ability to reduce the noise's interference on orientation and reconstruction.

\paragraph{\textbf{Models with Holes}} Models with holes are more complex and challenging to be dealt with, especially if the holes are narrow and deep. Although other methods have also paid attention to this issue and worked very hard to address it, they still encounter the problem of sealing holes incorrectly when facing small or deep holes.

\begin{figure*}
	\centering
	\includegraphics[width=1\linewidth]{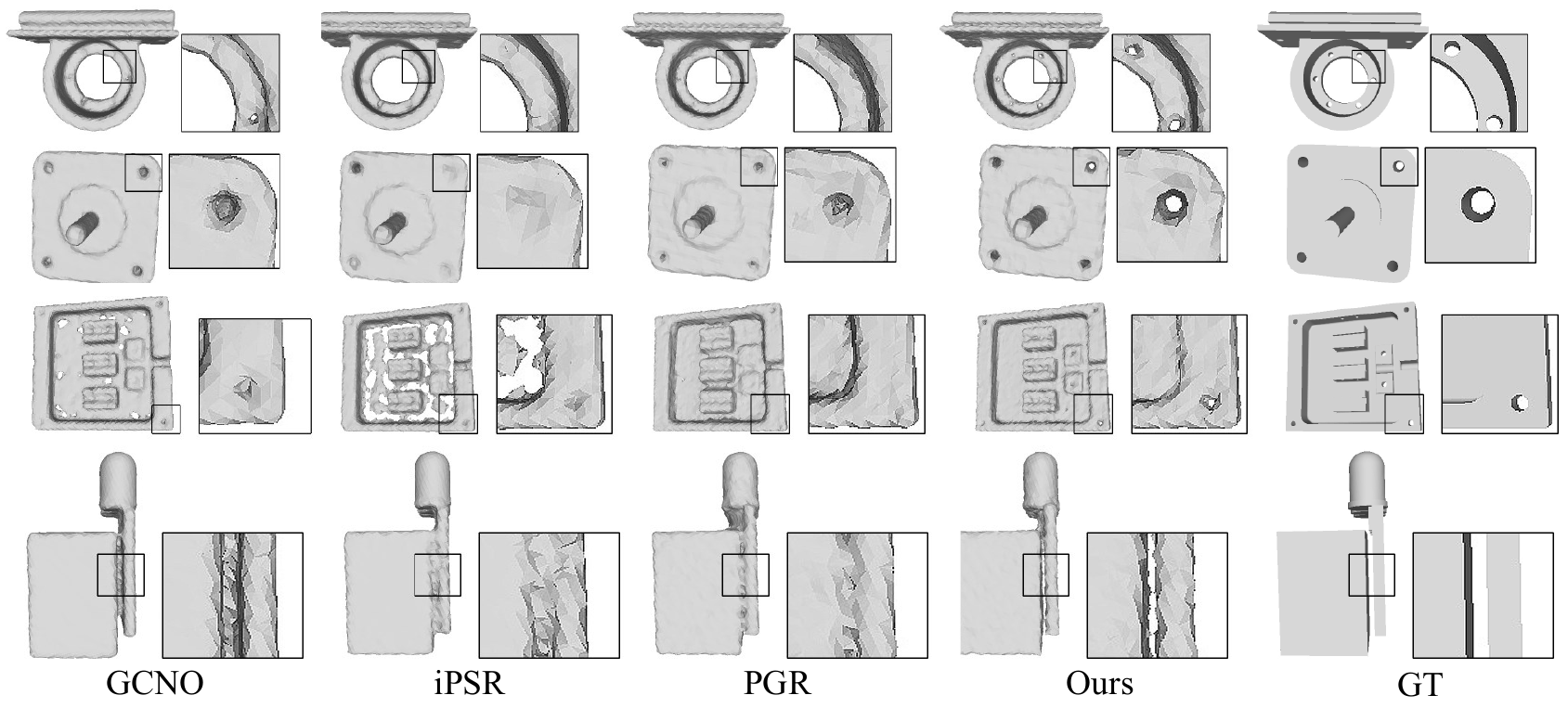}
	\caption{Qualitative comparison of our method with other state-of-the-art methods on reconstruction for point clouds with holes or fine seams. PGR cannot handle small holes (first row), deep holes (second row), thin structure combined with small holes (third row), and narrow, thin seams (fourth row), like iPSR and GCNO. By introducing anisotropy, our method can protect these valuable details and output the good reconstructions.}
	\label{holes}
\end{figure*}

\begin{figure*}
	\centering
	\includegraphics[width=1\linewidth]{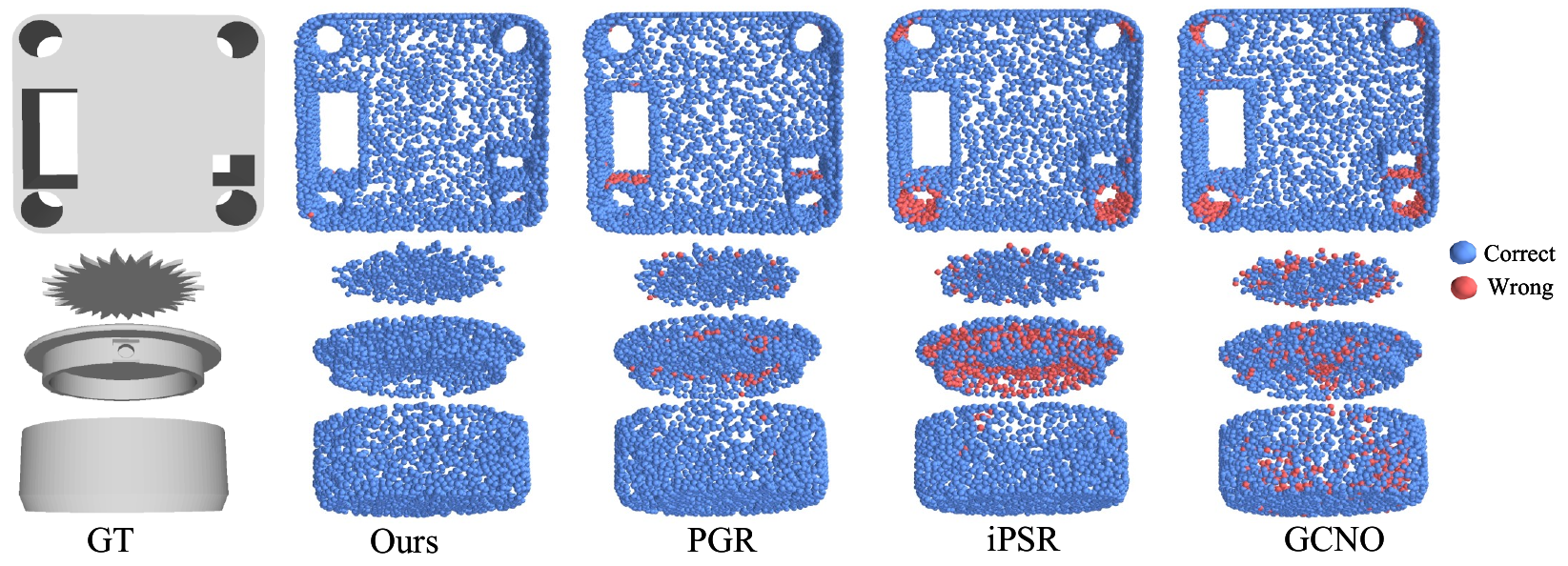}
	\caption{Qualitative comparison of orientation. Existing state-of-the-art orientation approaches cannot deal with various tricky examples such as thin structures, holes, and sharp features. Note that the red points indicate the wrong orientation. Our method further explores the potential of the Gauss formula for orientation and displays superior performances compared to these methods. }
	\label{orientation2}
\end{figure*}

Thanks to the introduction of anisotropy and the ability to establish more non-homogeneous equations with the proposed novel adaptive directional selection strategy, our method can protect these precious details and output high-quality orientation and reconstruction with state-of-the-art performance, while PGR cannot work well under the same parameter values. Figures \ref{holes} and \ref{orientation2} show a qualitative comparison of reconstruction and orientation.

\begin{figure*}[htbp]
	\centering
	\includegraphics[width=1\linewidth]{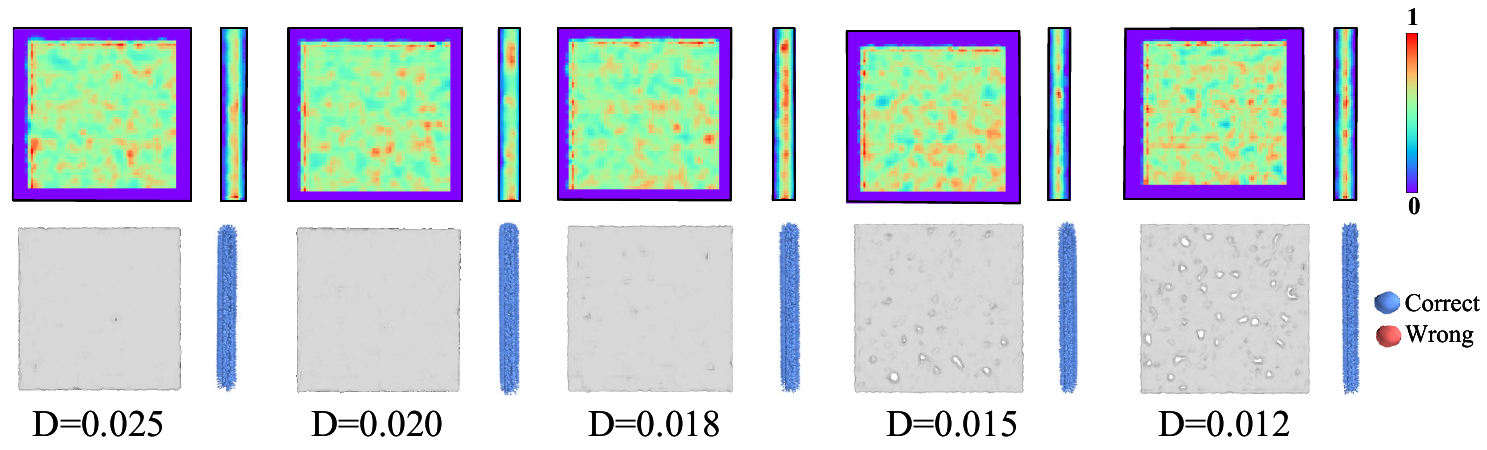}
	\caption{The reconstruction and orientation of our method on thin plates with thickness $D$ ranging from 0.012 to 0.025.}
	\label{thin2}
\end{figure*}

\paragraph{\textbf{Thin Structures}} As mentioned, thin structures often consist of two nearby surfaces with opposite normals. During normal estimation, the algorithm can be easily tricked into predicting the single surface's normal and aligning the orientation of one surface with the other, which leads to poor results in both orientation and reconstruction. However, our method introduces anisotropy and provides a novel adaptive selection strategy for directional information based on the analysis of the thin structures. Figure  \ref{thin2} showcases our reconstruction of thin plates with a length and width of 0.25 only with a thickness between 0.012 and 0.025. Figures \ref{thin1}, \ref{orientation1}, and  Table \ref{thin_plate_orientation} show the qualitative and quantitative comparisons of reconstruction results with other state-of-the-art methods. Our method can protect these valuable details with superior orientation and reconstruction for thin structures.

\begin{table}[htbp]
	\centering
	\caption{Quantitative comparisons of our method's orientation and reconstruction under different length parameter $L$  and regularization $\alpha$ on thin plates with a length and width of 0.5 and a thickness of 0.015 with other state-of-the-art methods under the same parameter values.}
	\begin{tabular}{|c|l|cccc|}
		\hline
		\multicolumn{2}{|c|}{} & $\text{PGP}_{90}$ $\uparrow $   & $\text{NC}_{p}$ $\uparrow $   & $\text{NC}_{s}$ $\uparrow $   & CD $\downarrow $ \\
		\hline
		\multicolumn{2}{|c|}{iPSR} & 0.9189 & 0.7853 & 0.7548 & 57.72 \\
		\cline{1-6}          
		\multirow{5}{*}{$\alpha$=2} & PGR   & 0.9786 & 0.9167 & 0.8569 & 17.89 \\
		& Ours(L=0.5) & 0.9834 & 0.9206 & 0.8743 & 8.75 \\
		& Ours(L=1) & 0.9835 & 0.9207 & 0.8740 & 8.66 \\
		& Ours(L=3) & \textbf{0.9848} & \textbf{0.9273} & 0.8784 & 8.09 \\
		& Ours(L=8) & 0.9832 & 0.9193 & 0.8785 & 6.75 \\
		\hline
		\multirow{5}{*}{$\alpha$=1.2} & PGR   & 0.9834 & 0.9168 & 0.8766 & 8.24 \\
		& Ours(L=0.5) & 0.9845 & 0.9206 & 0.8797 & 7.16 \\
		& Ours(L=1) & 0.9800 & 0.9185 & \textbf{0.8864} & \textbf{6.43} \\
		& Ours(L=3) & 0.9728 & 0.9054 & 0.8783 & 6.84 \\
		& Ours(L=8) & 0.9648 & 0.9037 & 0.8667 & 7.07 \\
		\hline
	\end{tabular}%
	\label{thin_plate_orientation}%
\end{table}%

\paragraph{\textbf{Sparse Point Clouds}}
Figure \ref{cond} shows the reconstruction of our method and PGR on a sparse point cloud with 500 points. Under the same parameter values, our method demonstrates superior reconstruction results compared to PGR, especially in continuity at the narrowed area.

\paragraph{\textbf{Running Speed and Storage Complexity}}

\begin{figure}
	\centering
	\includegraphics[width=1\linewidth]{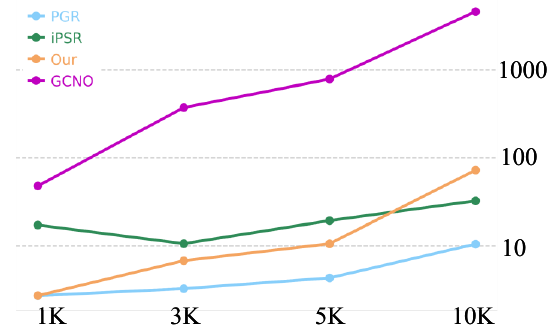}
	\caption{Qualitative and visualized comparison of algorithms' running time, where the y-coordinate uses logarithmic coordinates in seconds (/s). Our method  can maintain the good running speed on the whole.}
	\label{times}
\end{figure}

The running speed of algorithm is also an important indicator. Figure \ref{times} shows the time efficiency comparison of different reconstruction methods. Our method has a much faster computation speed than GCNO when handling point clouds ranging from 5K to 10K, and the overall time is similar to iPSR. Actually, For point clouds that are difficult to converge with, such as thin structures, the speed can be much faster than that of iPSR.

Nevertheless, we must admit that our method runs slower than PGR, although it is generally on the same order of magnitude. There are two main reasons for this: (1) we construct a larger number of equations, and (2) the analytical expression of the anisotropic fundamental solution and Guass kernel function in our method is more complex. This is also the area where we need to improve and optimize further. 

The comparison of running time between our method and other methods on point clouds of different scales is shown in Figure \ref{times}. Consistent with the previous theoretical analysis, our method requires a higher computational time than PGR due to the solving for the matrices with the size of $3N\times 3N$ instead of $N\times N$ in the experiment. Our method is slightly slower than iPSR with the same order of magnitude when dealing with dense point clouds and significantly faster than GCNO.

The storage of $B=A_{\bm{c}}A_{\bm{c}}^{T}+(\alpha-1)\cdot \textbf{diag}(A_{\bm{c}}A_{\bm{c}}^{T})$  is the square of the number of input samples. This is also our bottleneck. This makes it hard for our method to handle point clouds with very large scale, just like PGR. In the experimental setting, we considered the case of $m=3$. In contrast to PGR, which solves for the matrices with the size of $N \times N$ , our approach involves solving for the matrices with the size of $3N \times 3N$ . It is worth pointing out that although  the maximum capacity of our method is about 1/9 of PGR in theory, the novel blocking matrix strategy we proposed improves it to be 1/3.

At present, our method has a computational complexity of $O(N^{2})$ with the potential that the matrix-vector products $A_{\bm{c}} \xi$ and $A_{\bm{c}}^{T} \xi$ can be accelerated via the fast multipole method (\textbf{FMM}), ultimately reducing computational complexity to $O (N\text{log}N)$.

\subsection{Discussion of parameters}

\paragraph{\textbf{Module of Velocity Vectors}} With the proposed novel adaptive selection strategy for velocity vectors, our method controls the number of hyperparameters and only adds the length parameter $L$ to change the module of velocity vectors, compared to PGR. We conducted experiments to investigate the influence of length parameter  $L$ on normal estimation under different thicknesses $D$ (0.003 -0.008) of point clouds, as shown in Table \ref{Directions2}. Within a certain range, combined with the adaptive selection strategy, increasing the value of length parameter $L$ appropriately can improve the orientation quality of point clouds, especially for thin structures. Nevertheless, as the module of velocity vectors further increases, the corresponding truncation error will also increase due to the numerical approximation. Table \ref{thin_plate_orientation} also show the quantitative comparisons of our method's orientation and reconstruction under different length parameter $L$ on thin plates. The recommended range of length parameter $L$ is 0.5-6.

\begin{table}[htbp]\small
	\centering
	\caption{The quantitative comparison of length parameter  $L$ on normal estimation under different thicknesses $D$ (0.003 -0.008) of point clouds.}
	\begin{tabular}{c|ccccccc}
		\hline
		$\text{NC}_{p}$ $\uparrow $     & 0     & 0.5   & 1     & 2     & 4    & 6 & 8\\
		\hline
		0.008 & 0.8918 & 0.8970 & 0.9172 & 0.9194 & \textbf{0.9252} & 0.9248 & 0.9205\\
		0.005 & 0.8471 & 0.8782 & 0.8873 & 0.8905 & 0.9036 & \textbf{0.9103} & 0.9018\\
		0.004 & 0.8122 & 0.8564 & 0.8665 & 0.8739 & 0.8968 & \textbf{0.8976} & 0.8914\\
		0.003 & 0.7942 & 0.8291 & 0.8439 & 0.8624 & 0.8810 & \textbf{0.8881} & 0.8802\\
		\hline 
	\end{tabular}%
	\label{tab:addlabel}%
	\label{Directions2}
\end{table}%

\paragraph{\textbf{Regularization}}
PGR often suffers from its sensitivity to regularization terms. Overall, the too-small $\alpha$ can lead to artifacts and a rough appearance in reconstruction. Excessive $\alpha$ can cause the deviation between minimal-norm equation\eqref{4212} and the initial equation \eqref{4211}, resulting in two problems. The first is that the reconstructed surface is overly smooth, losing many details. The second is that the solved linear surface element leads to a decrease in the geometric meaning of the solution and the accuracy of orientation. Table \ref{thin_plate_orientation}  shows the orientation and reconstruction performance of our method and PGR under different levels of regularization. Thanks to anisotropy's introduction, our method can perform better than PGR at any regularization level, to a certain extent. Figures \ref{thin1},\ref{orientation1} show the orientation and reconstruction results of our method and PGR under the same regularization parameters. Our method has lower sensitivity and dependence on regularization. The qualitative comparisons are also shown in the supplementary materials.

\section{Limination}
\label{sec:Optimization}
However, our method still has shortcomings and room for improvement in the future. 

The first limitation lies in the memory and running time. Although we have proposed a blocking matrix strategy to save memory, the current bottleneck is the computation and storage of coefficient matrices like PGR. Due to the increase in the number of equations, the maximum capacity of our method is about 1/3 of PGR, approximately 15K. Although our method's speed is similar to iPSR and much faster than GCNO, it is slower than that of PGR due to establishing more equations, especially for the point clouds more than 10K.

The second limitation lies in the dependence and sensitivity of hyperparameters. Although we have significantly reduced the sensitivity and dependence on regularization terms $\alpha$ for solving the equations, our method still cannot completely eliminate the dependence on regularization terms. The qualitative comparisons are shown in the supplementary materials.

\begin{figure}
	\centering
	\includegraphics[width=1\linewidth]{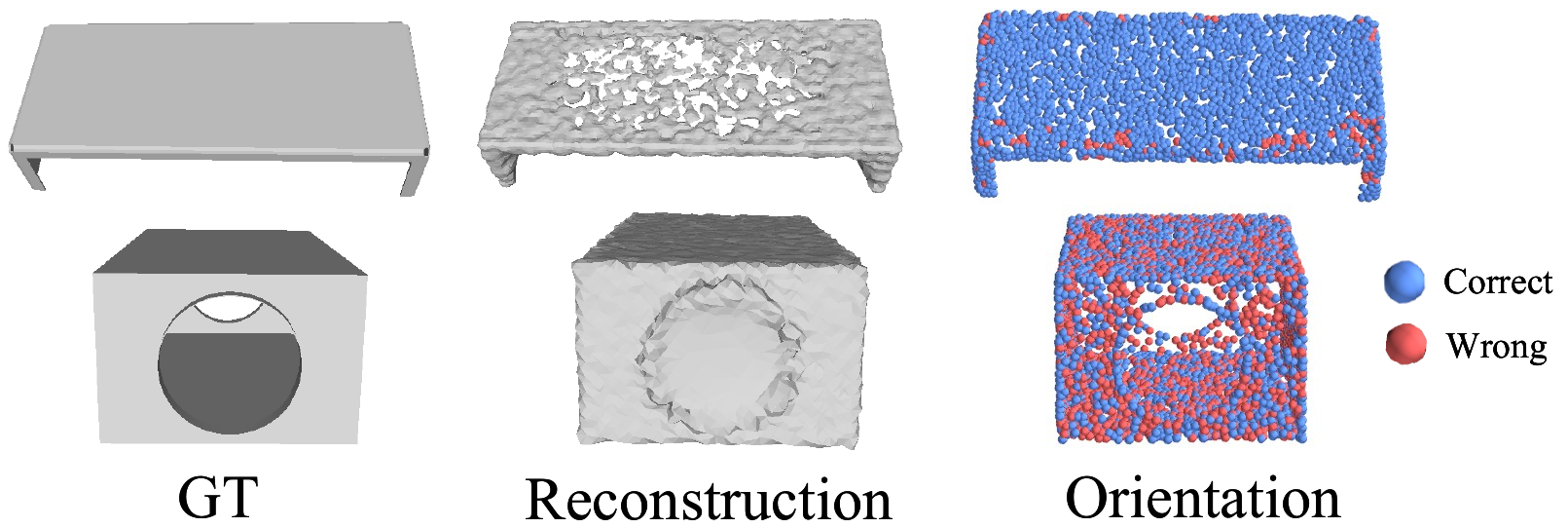}
	\caption{Limitations of our method. The first row shows that our method can orient correctly for some thin structures while reconstructing them poorly. The second row shows the thin structures with holes that our method cannot orient and reconstruct well, just like PGR and iPSR.}
	\label{Limination}
\end{figure}

In addition, introducing a convection term in the original Laplace operator makes the anisotropic  Gauss formula more complex. This puts higher requirements on the hyperparameter $w_{\text{min}}$, especially when the velocity vector's modulus is large. Excessive truncation may lead to a decrease in reconstruction performance. The truncation length $w_{\text{min}}$ is suggested to be set as 0.0015, which is also the default value of PGR.

In experiments, there are still some examples that our method cannot handle well. As shown in Figure \ref{Limination}, our method cannot deal with the orientation and reconstruction task of the sparse point clouds with thin structures and holes together well, just like other state-of-the-art methods. In addition, our method can improve the orientation for some difficult, thin structures but may show poor reconstruction performance. However, poor reconstruction is a common phenomenon. There still exists surface damage during reconstruction with the accurate values of the query points or taking the ground truth normals as the input.

\section{Future Work}
Our method still has many potentials worth exploring. In particular, the adaptive selection strategy for the velocity vector can be expanded in the following aspects.

\paragraph{\textbf{Weighting Factor}}We can further generalize the optimization problem \eqref{4223}  in the case of $ m \geqslant 3$, highlighting the information through weighting factors $\eta$. In detail,
\begin{align}
	\mathop{ \min }\limits_{\mu} \sum_{i=1}^{m}\eta_{i}||A_{\bm{c}_{i}}\mu-d_{1}||_{2}^{2},
\end{align}
which can be computed as
\begin{align*}
	H\mu=(\sum_{i=1}^{m}\eta_{i}A_{\bm{c}_{i}}^{T}d_{1}), \quad \text{with} \quad H=H_{0}+\text{diag}(H_{0}), H_{0}=(\sum_{i=1}^{m}\eta_{i}A_{\bm{c}_{i}}^{T}A_{\bm{c}_{i}}).
\end{align*}

It is an effective way to highlight the anisotropic information and improve the results. For example, we can use an adaptive parameter selection strategy combined with $\eta_{3}>1$ to enhance the constraint on the thinnest direction further.


\paragraph{\textbf{Point-wise Strategy}}We have proposed the adaptive strategy for velocity vectors by using the overall features of the point cloud, which has good performance on thin structures. However, it has not fully stimulated the potential of our method.

As shown in the anisotropic Gauss formula \eqref{GGL}, any velocity vector $\bm{c}$ can be used to calculate the indicator function. Compared to establishing a large number of equations through a unified velocity vector, we can tailor the velocity vector of each point in the anisotropic Gauss formula based on the characteristics of each input point cloud to utilize the local directional information.

The point cloud's local geometric features may not be consistent with the overall. Taking the surface composed of two perpendicular intersecting planes as an example, we tend to establish corresponding equations based on the correct normal of the point. However, using PCA on the whole point cloud can only estimate an overall normal. Hence, we propose a novel point-wise strategy by kNN and PCA at each point.

In detail, we further generalize $\bm{c}$ in equation \eqref{overall} to $\bm{c}_{\bm{p}_{i}}$,
\begin{align}
	A^{\bm{c}_{\bm{p}_{i}}}_{i}(\bm{p}_{i};\mathcal{P})\mu=\frac{1}{2},  \quad i=1,2,3,\cdots ,N ,
\end{align}
where $\bm{p}_{i} \in \mathcal{P}$ and $\bm{c}_{\bm{p}_{i}}$ can be obtained from the raw input cloud. 

If there is more detailed information about point clouds, such as being composed of two thin planes, we can take the points with similar local features as a whole, 
\begin{align}
	A^{\bm{c}_{\mathcal{P}_{i}}}_{i}(\mathcal{P}_{i};\mathcal{P})\mu=\frac{1}{2},
\end{align}
where $\mathcal{P}_{i} \subset \mathcal{P}$. The point-wise strategy can not only avoid the increase in running time and memory but also improve the quality of the equations and reduce the singularity of matrix $B$.

\section{Conclusion}
This paper showcases our new research effort towards extending the fundamental solution into an anisotropic form and the derivation of the corresponding anisotropic Gauss formula in theory. Compared to the reconstruction based on the original Gauss formula, our method can construct more non-homogeneous equations by further utilizing the information on point clouds and anisotropy.  

Furthermore, the introduction of anisotropy reduces our method's sensitivity to regularization parameters, improves the orientation, and enhances the algorithm's robustness. It significantly improves performance in the reconstruction of models with noise, holes, or thin structures. In response to the increasing number of equations, we propose numerical methods for solving under-determined and over-determined equation systems, respectively. 

In addition, we have explored the selection of velocity vectors deeply and proposed the adaptive selection strategy by PCA and SVD. For thin structures, our proposed strategy selects the eigenvectors of the point cloud, and further improves the orientation and reconstruction performance. 

Through extensive experiments, our approach shows superior performances in orientation and surface reconstruction compared to the other state-of-the-art methods.

\bibliographystyle{ACM-Reference-Format}
\normalem
\bibliography{sample-base}


\begin{thebibliography}{35}


\ifx \showCODEN    \undefined \def \showCODEN     #1{\unskip}     \fi
\ifx \showDOI      \undefined \def \showDOI       #1{#1}\fi
\ifx \showISBNx    \undefined \def \showISBNx     #1{\unskip}     \fi
\ifx \showISBNxiii \undefined \def \showISBNxiii  #1{\unskip}     \fi
\ifx \showISSN     \undefined \def \showISSN      #1{\unskip}     \fi
\ifx \showLCCN     \undefined \def \showLCCN      #1{\unskip}     \fi
\ifx \shownote     \undefined \def \shownote      #1{#1}          \fi
\ifx \showarticletitle \undefined \def \showarticletitle #1{#1}   \fi
\ifx \showURL      \undefined \def \showURL       {\relax}        \fi
\providecommand\bibfield[2]{#2}
\providecommand\bibinfo[2]{#2}
\providecommand\natexlab[1]{#1}
\providecommand\showeprint[2][]{arXiv:#2}

\bibitem[\protect\citeauthoryear{Ben-Shabat, Koneputugodage, and
  Gould}{Ben-Shabat et~al\mbox{.}}{2022}]%
        {ben2022digs}
\bibfield{author}{\bibinfo{person}{Yizhak Ben-Shabat},
  \bibinfo{person}{Chamin~Hewa Koneputugodage}, {and} \bibinfo{person}{Stephen
  Gould}.} \bibinfo{year}{2022}\natexlab{}.
\newblock \showarticletitle{Digs: Divergence guided shape implicit neural
  representation for unoriented point clouds}. In
  \bibinfo{booktitle}{\emph{Proceedings of the IEEE/CVF Conference on Computer
  Vision and Pattern Recognition}}. \bibinfo{pages}{19323--19332}.
\newblock


\bibitem[\protect\citeauthoryear{Calakli and Taubin}{Calakli and
  Taubin}{2011}]%
        {calakli2011ssd}
\bibfield{author}{\bibinfo{person}{Fatih Calakli} {and}
  \bibinfo{person}{Gabriel Taubin}.} \bibinfo{year}{2011}\natexlab{}.
\newblock \showarticletitle{S{SD}: Smooth signed distance surface
  reconstruction}. In \bibinfo{booktitle}{\emph{Computer Graphics Forum}},
  Vol.~\bibinfo{volume}{30}. Wiley Online Library, \bibinfo{pages}{1993--2002}.
\newblock


\bibitem[\protect\citeauthoryear{Darve}{Darve}{2000}]%
        {DARVE2000195}
\bibfield{author}{\bibinfo{person}{Eric Darve}.}
  \bibinfo{year}{2000}\natexlab{}.
\newblock \showarticletitle{The Fast Multipole Method: Numerical
  Implementation}.
\newblock \bibinfo{journal}{\emph{J. Comput. Phys.}} \bibinfo{volume}{160},
  \bibinfo{number}{1} (\bibinfo{year}{2000}), \bibinfo{pages}{195--240}.
\newblock
\showISSN{0021-9991}
\urldef\tempurl%
\url{https://doi.org/10.1006/jcph.2000.6451}
\showDOI{\tempurl}


\bibitem[\protect\citeauthoryear{Dey and Sun}{Dey and Sun}{2005}]%
        {dey2005adaptive}
\bibfield{author}{\bibinfo{person}{Tamal~K Dey} {and} \bibinfo{person}{Jian
  Sun}.} \bibinfo{year}{2005}\natexlab{}.
\newblock \showarticletitle{. An Adaptive MLS Surface for Reconstruction with
  Guarantees.}. In \bibinfo{booktitle}{\emph{Symposium on Geometry
  processing}}. \bibinfo{pages}{43--52}.
\newblock


\bibitem[\protect\citeauthoryear{Erler, Guerrero, Ohrhallinger, Mitra, and
  Wimmer}{Erler et~al\mbox{.}}{2020}]%
        {erler2020points2surf}
\bibfield{author}{\bibinfo{person}{Philipp Erler}, \bibinfo{person}{Paul
  Guerrero}, \bibinfo{person}{Stefan Ohrhallinger}, \bibinfo{person}{Niloy~J
  Mitra}, {and} \bibinfo{person}{Michael Wimmer}.}
  \bibinfo{year}{2020}\natexlab{}.
\newblock \showarticletitle{Points2surf learning implicit surfaces from point
  clouds}. In \bibinfo{booktitle}{\emph{European Conference on Computer
  Vision}}. Springer, \bibinfo{pages}{108--124}.
\newblock


\bibitem[\protect\citeauthoryear{Gropp, Yariv, Haim, Atzmon, and Lipman}{Gropp
  et~al\mbox{.}}{2020}]%
        {gropp2020implicit}
\bibfield{author}{\bibinfo{person}{Amos Gropp}, \bibinfo{person}{Lior Yariv},
  \bibinfo{person}{Niv Haim}, \bibinfo{person}{Matan Atzmon}, {and}
  \bibinfo{person}{Yaron Lipman}.} \bibinfo{year}{2020}\natexlab{}.
\newblock \showarticletitle{Implicit geometric regularization for learning
  shapes}.
\newblock \bibinfo{journal}{\emph{arXiv preprint arXiv:2002.10099}}
  (\bibinfo{year}{2020}).
\newblock


\bibitem[\protect\citeauthoryear{Guerrero, Kleiman, Ovsjanikov, and
  Mitra}{Guerrero et~al\mbox{.}}{2018}]%
        {guerrero2018pcpnet}
\bibfield{author}{\bibinfo{person}{Paul Guerrero}, \bibinfo{person}{Yanir
  Kleiman}, \bibinfo{person}{Maks Ovsjanikov}, {and} \bibinfo{person}{Niloy~J
  Mitra}.} \bibinfo{year}{2018}\natexlab{}.
\newblock \showarticletitle{Pcpnet learning local shape properties from raw
  point clouds}. In \bibinfo{booktitle}{\emph{Computer graphics forum}},
  Vol.~\bibinfo{volume}{37}. Wiley Online Library, \bibinfo{pages}{75--85}.
\newblock


\bibitem[\protect\citeauthoryear{Hoppe, DeRose, Duchamp, McDonald, and
  Stuetzle}{Hoppe et~al\mbox{.}}{1992}]%
        {hoppe1992surface}
\bibfield{author}{\bibinfo{person}{Hugues Hoppe}, \bibinfo{person}{Tony
  DeRose}, \bibinfo{person}{Tom Duchamp}, \bibinfo{person}{John McDonald},
  {and} \bibinfo{person}{Werner Stuetzle}.} \bibinfo{year}{1992}\natexlab{}.
\newblock \showarticletitle{Surface reconstruction from unorganized points}. In
  \bibinfo{booktitle}{\emph{Proceedings of the 19th annual conference on
  computer graphics and interactive techniques}}. \bibinfo{pages}{71--78}.
\newblock


\bibitem[\protect\citeauthoryear{Hou, Wang, Wang, Qin, Qian, and He}{Hou
  et~al\mbox{.}}{2022}]%
        {hou2022iterative}
\bibfield{author}{\bibinfo{person}{Fei Hou}, \bibinfo{person}{Chiyu Wang},
  \bibinfo{person}{Wencheng Wang}, \bibinfo{person}{Hong Qin},
  \bibinfo{person}{Chen Qian}, {and} \bibinfo{person}{Ying He}.}
  \bibinfo{year}{2022}\natexlab{}.
\newblock \showarticletitle{Iterative poisson surface reconstruction (ipsr) for
  unoriented points}.
\newblock \bibinfo{journal}{\emph{arXiv preprint arXiv:2209.09510}}
  (\bibinfo{year}{2022}).
\newblock


\bibitem[\protect\citeauthoryear{Huang, Carr, and Ju}{Huang
  et~al\mbox{.}}{2019}]%
        {huang2019variational}
\bibfield{author}{\bibinfo{person}{Zhiyang Huang}, \bibinfo{person}{Nathan
  Carr}, {and} \bibinfo{person}{Tao Ju}.} \bibinfo{year}{2019}\natexlab{}.
\newblock \showarticletitle{Variational implicit point set surfaces}.
\newblock \bibinfo{journal}{\emph{ACM Transactions on Graphics (TOG)}}
  \bibinfo{volume}{38}, \bibinfo{number}{4} (\bibinfo{year}{2019}),
  \bibinfo{pages}{1--13}.
\newblock


\bibitem[\protect\citeauthoryear{Ijiri, Yoshizawa, Sato, Ito, and Yokota}{Ijiri
  et~al\mbox{.}}{2013}]%
        {ijiri2013bilateral}
\bibfield{author}{\bibinfo{person}{Takashi Ijiri}, \bibinfo{person}{Shin
  Yoshizawa}, \bibinfo{person}{Yu Sato}, \bibinfo{person}{Masaaki Ito}, {and}
  \bibinfo{person}{Hideo Yokota}.} \bibinfo{year}{2013}\natexlab{}.
\newblock \showarticletitle{Bilateral Hermite Radial Basis Functions for
  Contour-based Volume Segmentation}. In \bibinfo{booktitle}{\emph{Computer
  Graphics Forum}}, Vol.~\bibinfo{volume}{32}. Wiley Online Library,
  \bibinfo{pages}{123--132}.
\newblock


\bibitem[\protect\citeauthoryear{Kazhdan}{Kazhdan}{2005}]%
        {kazhdan2005reconstruction}
\bibfield{author}{\bibinfo{person}{Michael Kazhdan}.}
  \bibinfo{year}{2005}\natexlab{}.
\newblock \showarticletitle{Reconstruction of solid models from oriented point
  sets}. In \bibinfo{booktitle}{\emph{Proceedings of the third Eurographics
  symposium on Geometry processing}}. \bibinfo{pages}{73--es}.
\newblock


\bibitem[\protect\citeauthoryear{Kazhdan, Bolitho, and Hoppe}{Kazhdan
  et~al\mbox{.}}{2006}]%
        {kazhdan2006poisson}
\bibfield{author}{\bibinfo{person}{Michael Kazhdan}, \bibinfo{person}{Matthew
  Bolitho}, {and} \bibinfo{person}{Hugues Hoppe}.}
  \bibinfo{year}{2006}\natexlab{}.
\newblock \showarticletitle{Poisson surface reconstruction}. In
  \bibinfo{booktitle}{\emph{Proceedings of the fourth Eurographics symposium on
  Geometry processing}}, Vol.~\bibinfo{volume}{7}. \bibinfo{pages}{0}.
\newblock


\bibitem[\protect\citeauthoryear{Kazhdan, Chuang, Rusinkiewicz, and
  Hoppe}{Kazhdan et~al\mbox{.}}{2020}]%
        {kazhdan2020poisson}
\bibfield{author}{\bibinfo{person}{Misha Kazhdan}, \bibinfo{person}{Ming
  Chuang}, \bibinfo{person}{Szymon Rusinkiewicz}, {and} \bibinfo{person}{Hugues
  Hoppe}.} \bibinfo{year}{2020}\natexlab{}.
\newblock \showarticletitle{Poisson surface reconstruction with envelope
  constraints}. In \bibinfo{booktitle}{\emph{Computer graphics forum}},
  Vol.~\bibinfo{volume}{39}. Wiley Online Library, \bibinfo{pages}{173--182}.
\newblock


\bibitem[\protect\citeauthoryear{Kazhdan and Hoppe}{Kazhdan and Hoppe}{2013}]%
        {kazhdan2013screened}
\bibfield{author}{\bibinfo{person}{Michael Kazhdan} {and}
  \bibinfo{person}{Hugues Hoppe}.} \bibinfo{year}{2013}\natexlab{}.
\newblock \showarticletitle{Screened poisson surface reconstruction}.
\newblock \bibinfo{journal}{\emph{ACM Transactions on Graphics (ToG)}}
  \bibinfo{volume}{32}, \bibinfo{number}{3} (\bibinfo{year}{2013}),
  \bibinfo{pages}{1--13}.
\newblock


\bibitem[\protect\citeauthoryear{Kingma and Ba}{Kingma and Ba}{2014}]%
        {kingma2014adam}
\bibfield{author}{\bibinfo{person}{Diederik~P Kingma} {and}
  \bibinfo{person}{Jimmy Ba}.} \bibinfo{year}{2014}\natexlab{}.
\newblock \showarticletitle{Adam: A method for stochastic optimization}.
\newblock \bibinfo{journal}{\emph{arXiv preprint arXiv:1412.6980}}
  (\bibinfo{year}{2014}).
\newblock


\bibitem[\protect\citeauthoryear{Kolluri}{Kolluri}{2008}]%
        {kolluri2008provably}
\bibfield{author}{\bibinfo{person}{Ravikrishna Kolluri}.}
  \bibinfo{year}{2008}\natexlab{}.
\newblock \showarticletitle{Provably good moving least squares}.
\newblock \bibinfo{journal}{\emph{ACM Transactions on Algorithms (TALG)}}
  \bibinfo{volume}{4}, \bibinfo{number}{2} (\bibinfo{year}{2008}),
  \bibinfo{pages}{1--25}.
\newblock


\bibitem[\protect\citeauthoryear{Li, Feng, Shi, Gao, Fang, Liu, and Han}{Li
  et~al\mbox{.}}{2023}]%
        {li2023shs}
\bibfield{author}{\bibinfo{person}{Qing Li}, \bibinfo{person}{Huifang Feng},
  \bibinfo{person}{Kanle Shi}, \bibinfo{person}{Yue Gao}, \bibinfo{person}{Yi
  Fang}, \bibinfo{person}{Yu-Shen Liu}, {and} \bibinfo{person}{Zhizhong Han}.}
  \bibinfo{year}{2023}\natexlab{}.
\newblock \showarticletitle{Shs-net: Learning signed hyper surfaces for
  oriented normal estimation of point clouds}. In
  \bibinfo{booktitle}{\emph{Proceedings of the IEEE/CVF Conference on Computer
  Vision and Pattern Recognition}}. \bibinfo{pages}{13591--13600}.
\newblock


\bibitem[\protect\citeauthoryear{Lin, Xiao, Shi, and Wang}{Lin
  et~al\mbox{.}}{2023}]%
        {2022PGR}
\bibfield{author}{\bibinfo{person}{Siyou Lin}, \bibinfo{person}{Dong Xiao},
  \bibinfo{person}{Zuoqiang Shi}, {and} \bibinfo{person}{Bin Wang}.}
  \bibinfo{year}{2023}\natexlab{}.
\newblock \showarticletitle{Surface Reconstruction from Point Clouds without
  Normals by Parametrizing the Gauss Formula}.
\newblock \bibinfo{journal}{\emph{{ACM} Trans. Graph.}} \bibinfo{volume}{42},
  \bibinfo{number}{2} (\bibinfo{year}{2023}), \bibinfo{pages}{14:1--14:19}.
\newblock


\bibitem[\protect\citeauthoryear{Liu, Wang, Brunnett, and Wang}{Liu
  et~al\mbox{.}}{2016}]%
        {liu2016closed}
\bibfield{author}{\bibinfo{person}{Shengjun Liu}, \bibinfo{person}{Charlie~CL
  Wang}, \bibinfo{person}{Guido Brunnett}, {and} \bibinfo{person}{Jun Wang}.}
  \bibinfo{year}{2016}\natexlab{}.
\newblock \showarticletitle{A closed-form formulation of HRBF-based surface
  reconstruction by approximate solution}.
\newblock \bibinfo{journal}{\emph{Computer-Aided Design}}  \bibinfo{volume}{78}
  (\bibinfo{year}{2016}), \bibinfo{pages}{147--157}.
\newblock


\bibitem[\protect\citeauthoryear{Lu, Shi, Sun, and Wang}{Lu
  et~al\mbox{.}}{2019}]%
        {2019GR}
\bibfield{author}{\bibinfo{person}{Wenjia Lu}, \bibinfo{person}{Zuoqiang Shi},
  \bibinfo{person}{Jian Sun}, {and} \bibinfo{person}{Bin Wang}.}
  \bibinfo{year}{2019}\natexlab{}.
\newblock \showarticletitle{Surface Reconstruction Based on the Modified Gauss
  Formula}.
\newblock \bibinfo{journal}{\emph{{ACM} Trans. Graph.}} \bibinfo{volume}{38},
  \bibinfo{number}{1} (\bibinfo{year}{2019}), \bibinfo{pages}{2:1--2:18}.
\newblock


\bibitem[\protect\citeauthoryear{Mac{\^e}do, Gois, and Velho}{Mac{\^e}do
  et~al\mbox{.}}{2011}]%
        {macedo2011hermite}
\bibfield{author}{\bibinfo{person}{Ives Mac{\^e}do},
  \bibinfo{person}{Joao~Paulo Gois}, {and} \bibinfo{person}{Luiz Velho}.}
  \bibinfo{year}{2011}\natexlab{}.
\newblock \showarticletitle{Hermite radial basis functions implicits}. In
  \bibinfo{booktitle}{\emph{Computer Graphics Forum}},
  Vol.~\bibinfo{volume}{30}. Wiley Online Library, \bibinfo{pages}{27--42}.
\newblock


\bibitem[\protect\citeauthoryear{Manson, Petrova, and Schaefer}{Manson
  et~al\mbox{.}}{2008}]%
        {manson2008streaming}
\bibfield{author}{\bibinfo{person}{Josiah Manson}, \bibinfo{person}{Guergana
  Petrova}, {and} \bibinfo{person}{Scott Schaefer}.}
  \bibinfo{year}{2008}\natexlab{}.
\newblock \showarticletitle{Streaming surface reconstruction using wavelets}.
  In \bibinfo{booktitle}{\emph{Computer Graphics Forum}},
  Vol.~\bibinfo{volume}{27}. Wiley Online Library, \bibinfo{pages}{1411--1420}.
\newblock


\bibitem[\protect\citeauthoryear{McIntyre and Cairns}{McIntyre and
  Cairns}{1993}]%
        {mcintyre1993new}
\bibfield{author}{\bibinfo{person}{Margaret McIntyre} {and}
  \bibinfo{person}{Grant Cairns}.} \bibinfo{year}{1993}\natexlab{}.
\newblock \showarticletitle{A new formula for winding number}.
\newblock \bibinfo{journal}{\emph{Geometriae Dedicata}} \bibinfo{volume}{46},
  \bibinfo{number}{2} (\bibinfo{year}{1993}), \bibinfo{pages}{149--159}.
\newblock


\bibitem[\protect\citeauthoryear{Metzer, Hanocka, Zorin, Giryes, Panozzo, and
  Cohen-Or}{Metzer et~al\mbox{.}}{2021}]%
        {metzer2021orienting}
\bibfield{author}{\bibinfo{person}{Gal Metzer}, \bibinfo{person}{Rana Hanocka},
  \bibinfo{person}{Denis Zorin}, \bibinfo{person}{Raja Giryes},
  \bibinfo{person}{Daniele Panozzo}, {and} \bibinfo{person}{Daniel Cohen-Or}.}
  \bibinfo{year}{2021}\natexlab{}.
\newblock \showarticletitle{Orienting point clouds with dipole propagation}.
\newblock \bibinfo{journal}{\emph{ACM Transactions on Graphics (TOG)}}
  \bibinfo{volume}{40}, \bibinfo{number}{4} (\bibinfo{year}{2021}),
  \bibinfo{pages}{1--14}.
\newblock


\bibitem[\protect\citeauthoryear{Morse, Yoo, Rheingans, Chen, and
  Subramanian}{Morse et~al\mbox{.}}{2001}]%
        {923379}
\bibfield{author}{\bibinfo{person}{B.S. Morse}, \bibinfo{person}{T.S. Yoo},
  \bibinfo{person}{P. Rheingans}, \bibinfo{person}{D.T. Chen}, {and}
  \bibinfo{person}{K.R. Subramanian}.} \bibinfo{year}{2001}\natexlab{}.
\newblock \showarticletitle{Interpolating implicit surfaces from scattered
  surface data using compactly supported radial basis functions}. In
  \bibinfo{booktitle}{\emph{Proceedings International Conference on Shape
  Modeling and Applications}}. \bibinfo{pages}{89--98}.
\newblock
\urldef\tempurl%
\url{https://doi.org/10.1109/SMA.2001.923379}
\showDOI{\tempurl}


\bibitem[\protect\citeauthoryear{Nishino and Loomis}{Nishino and
  Loomis}{2017}]%
        {nishino2017cupy}
\bibfield{author}{\bibinfo{person}{ROYUD Nishino} {and} \bibinfo{person}{Shohei
  Hido~Crissman Loomis}.} \bibinfo{year}{2017}\natexlab{}.
\newblock \showarticletitle{Cupy: A numpy-compatible library for nvidia gpu
  calculations}.
\newblock \bibinfo{journal}{\emph{31st confernce on neural information
  processing systems}} \bibinfo{volume}{151}, \bibinfo{number}{7}
  (\bibinfo{year}{2017}).
\newblock


\bibitem[\protect\citeauthoryear{Peng, Jiang, Liao, Niemeyer, Pollefeys, and
  Geiger}{Peng et~al\mbox{.}}{2021}]%
        {peng2021shape}
\bibfield{author}{\bibinfo{person}{Songyou Peng}, \bibinfo{person}{Chiyu
  Jiang}, \bibinfo{person}{Yiyi Liao}, \bibinfo{person}{Michael Niemeyer},
  \bibinfo{person}{Marc Pollefeys}, {and} \bibinfo{person}{Andreas Geiger}.}
  \bibinfo{year}{2021}\natexlab{}.
\newblock \showarticletitle{Shape as points: A differentiable poisson solver}.
\newblock \bibinfo{journal}{\emph{Advances in Neural Information Processing
  Systems}}  \bibinfo{volume}{34} (\bibinfo{year}{2021}),
  \bibinfo{pages}{13032--13044}.
\newblock


\bibitem[\protect\citeauthoryear{Ren, Lyu, He, Cao, Yang, Sheng, Zhang, and
  Wu}{Ren et~al\mbox{.}}{2018}]%
        {ren2018biorthogonal}
\bibfield{author}{\bibinfo{person}{Xiaohua Ren}, \bibinfo{person}{Luan Lyu},
  \bibinfo{person}{Xiaowei He}, \bibinfo{person}{Wei Cao},
  \bibinfo{person}{Zhixin Yang}, \bibinfo{person}{Bin Sheng},
  \bibinfo{person}{Yanci Zhang}, {and} \bibinfo{person}{Enhua Wu}.}
  \bibinfo{year}{2018}\natexlab{}.
\newblock \showarticletitle{Biorthogonal wavelet surface reconstruction using
  partial integrations}. In \bibinfo{booktitle}{\emph{Computer Graphics
  Forum}}, Vol.~\bibinfo{volume}{37}. Wiley Online Library,
  \bibinfo{pages}{13--24}.
\newblock


\bibitem[\protect\citeauthoryear{Walder, Chapelle, and Sch{\"o}lkopf}{Walder
  et~al\mbox{.}}{2006}]%
        {walder2006implicit}
\bibfield{author}{\bibinfo{person}{Christian Walder}, \bibinfo{person}{Olivier
  Chapelle}, {and} \bibinfo{person}{Bernhard Sch{\"o}lkopf}.}
  \bibinfo{year}{2006}\natexlab{}.
\newblock \showarticletitle{Implicit surfaces with globally regularised and
  compactly supported basis functions}.
\newblock \bibinfo{journal}{\emph{Advances in Neural Information Processing
  Systems}}  \bibinfo{volume}{19} (\bibinfo{year}{2006}).
\newblock


\bibitem[\protect\citeauthoryear{Wang, Zhang, Xu, Zhang, Wang, Chen, Xin, Wang,
  and Tu}{Wang et~al\mbox{.}}{2023}]%
        {wang2023neural}
\bibfield{author}{\bibinfo{person}{Zixiong Wang}, \bibinfo{person}{Yunxiao
  Zhang}, \bibinfo{person}{Rui Xu}, \bibinfo{person}{Fan Zhang},
  \bibinfo{person}{Peng-Shuai Wang}, \bibinfo{person}{Shuangmin Chen},
  \bibinfo{person}{Shiqing Xin}, \bibinfo{person}{Wenping Wang}, {and}
  \bibinfo{person}{Changhe Tu}.} \bibinfo{year}{2023}\natexlab{}.
\newblock \showarticletitle{Neural-Singular-Hessian: Implicit Neural
  Representation of Unoriented Point Clouds by Enforcing Singular Hessian}.
\newblock \bibinfo{journal}{\emph{ACM Transactions on Graphics (TOG)}}
  \bibinfo{volume}{42}, \bibinfo{number}{6} (\bibinfo{year}{2023}),
  \bibinfo{pages}{1--14}.
\newblock


\bibitem[\protect\citeauthoryear{Xiao, Shi, Li, Deng, and Wang}{Xiao
  et~al\mbox{.}}{2023}]%
        {xiao2023point}
\bibfield{author}{\bibinfo{person}{Dong Xiao}, \bibinfo{person}{Zuoqiang Shi},
  \bibinfo{person}{Siyu Li}, \bibinfo{person}{Bailin Deng}, {and}
  \bibinfo{person}{Bin Wang}.} \bibinfo{year}{2023}\natexlab{}.
\newblock \showarticletitle{Point normal orientation and surface reconstruction
  by incorporating isovalue constraints to Poisson equation}.
\newblock \bibinfo{journal}{\emph{Computer Aided Geometric Design}}
  \bibinfo{volume}{103} (\bibinfo{year}{2023}), \bibinfo{pages}{102195}.
\newblock


\bibitem[\protect\citeauthoryear{Xu, Dou, Wang, Xin, Chen, Jiang, Guo, Wang,
  and Tu}{Xu et~al\mbox{.}}{2023a}]%
        {2023GCNO}
\bibfield{author}{\bibinfo{person}{Rui Xu}, \bibinfo{person}{Zhiyang Dou},
  \bibinfo{person}{Ningna Wang}, \bibinfo{person}{Shiqing Xin},
  \bibinfo{person}{Shuang{-}Min Chen}, \bibinfo{person}{Mingyan Jiang},
  \bibinfo{person}{Xiaohu Guo}, \bibinfo{person}{Wenping Wang}, {and}
  \bibinfo{person}{Changhe Tu}.} \bibinfo{year}{2023}\natexlab{a}.
\newblock \showarticletitle{Globally Consistent Normal Orientation for Point
  Clouds by Regularizing the Winding-Number Field}.
\newblock \bibinfo{journal}{\emph{{ACM} Trans. Graph.}} \bibinfo{volume}{42},
  \bibinfo{number}{4} (\bibinfo{year}{2023}), \bibinfo{pages}{111:1--111:15}.
\newblock


\bibitem[\protect\citeauthoryear{Xu, Dou, Wang, Xin, Chen, Jiang, Guo, Wang,
  and Tu}{Xu et~al\mbox{.}}{2023b}]%
        {xu2023globally}
\bibfield{author}{\bibinfo{person}{Rui Xu}, \bibinfo{person}{Zhiyang Dou},
  \bibinfo{person}{Ningna Wang}, \bibinfo{person}{Shiqing Xin},
  \bibinfo{person}{Shuangmin Chen}, \bibinfo{person}{Mingyan Jiang},
  \bibinfo{person}{Xiaohu Guo}, \bibinfo{person}{Wenping Wang}, {and}
  \bibinfo{person}{Changhe Tu}.} \bibinfo{year}{2023}\natexlab{b}.
\newblock \showarticletitle{Globally consistent normal orientation for point
  clouds by regularizing the winding-number field}.
\newblock \bibinfo{journal}{\emph{ACM Transactions on Graphics (TOG)}}
  \bibinfo{volume}{42}, \bibinfo{number}{4} (\bibinfo{year}{2023}),
  \bibinfo{pages}{1--15}.
\newblock


\bibitem[\protect\citeauthoryear{Zhou, Ma, Liu, Fang, and Han}{Zhou
  et~al\mbox{.}}{2022}]%
        {zhou2022learning}
\bibfield{author}{\bibinfo{person}{Junsheng Zhou}, \bibinfo{person}{Baorui Ma},
  \bibinfo{person}{Yu-Shen Liu}, \bibinfo{person}{Yi Fang}, {and}
  \bibinfo{person}{Zhizhong Han}.} \bibinfo{year}{2022}\natexlab{}.
\newblock \showarticletitle{Learning consistency-aware unsigned distance
  functions progressively from raw point clouds}.
\newblock \bibinfo{journal}{\emph{Advances in Neural Information Processing
  Systems}}  \bibinfo{volume}{35} (\bibinfo{year}{2022}),
  \bibinfo{pages}{16481--16494}.
\newblock


\end{thebibliography}
\appendix
\section{ derivation of isotropic fundamental solutions}\label{A}
The fundamental solution of $n-$dimensions has the generalized analytical formula
\begin{equation}\label{E1}
\Phi(\bm{x})=\left\{
\begin{aligned}
&-\dfrac{1}{2\pi} \text{log} | \bm{x} | ,& &  n=2\\
&\dfrac{1}{n(n-2)\omega(n)} \dfrac{1}{|\bm{x}|^{n-2}} ,& &    n \geqslant 3
\end{aligned} \right.
\end{equation}
of the original Laplace equation  $\Delta u = 0 $. Where $\omega(n)$ is the volume of a n-dimensional unit sphere, namely $\omega(n)=\pi^{\frac{n}{2}}/\Gamma(\frac{n}{2}+1)$, and $\Gamma(\cdot)$ is the Gamma function. 

  Firstly, we utilize symmetry to search for its radial solution, treating it as a function of $ r=|\bm{x}|$, which is only related to the length of the $\bm{x}$. Suppose that $u(\bm{x})=v(r),\bm{x}=(x_{1},x_{2},x_{3},\cdots,x_{n})$, then 
\begin{align*}
    u_{x_{i}}=v'(r)\dfrac{x_{i}}{r},u_{x_{i}x_{i}}=v''(r)\dfrac{x_{i}^{2}}{r^{2}}+v'(r)(\dfrac{1}{r}-\dfrac{x_{i}^{2}}{r^{3}}), \quad i=1,2,\cdots n.
\end{align*}

Thus
\begin{align*}
\Delta u&= v''(r)\dfrac{x_{1}^{2}+x_{2}^{2}+x_{3}^{2}+\cdots+x_{n}^{2}}{r^{2}}+v'(r)(\dfrac{n}{r}-\dfrac{x_{1}^{2}+x_{2}^{2}+x_{3}^{2}+\cdots+x_{n}^{2}}{r^{3}})\\
&=v''(r)+v'(r)(\dfrac{n-1}{r}).
\end{align*}

Due to $\Delta u=0$, then 
\begin{align*}
v''(r)+v'(r)(\frac{n-1}{r})=0. 
\end{align*}

This is a second-order ordinary differential equation (ODE) about scalar $r$. When $n \geqslant 3$, it can be solved by quadratic integration. We can also solve it by using the method of constant variation, assuming that it has the form of $v=r^{m}$.    

\section{ Derivation of Anisotropic fundamental solutions}\label{B}
This is the calculation process of Equation \eqref{GFF2}. 

Due to the addition of velocity vector $\bm{c}$, the generalized Laplace equation \eqref{GL2} may not have symmetry, so it cannot be simply processed according to the previous method of solving isotropic equations. Firstly, let 
\begin{align*}
   u(\bm{x})=v(\bm{x})e^{\frac{1}{2}\bm{c}\cdot \bm{x}}, 
\end{align*}
then 
\begin{align*}
    \nabla u(\bm{x})&=\nabla(v(\bm{x})e^{\frac{1}{2}\bm{c}\cdot \bm{x}})\\
    &=e^{\frac{1}{2}\bm{c}\cdot \bm{x}}  \nabla v(\bm{x})+ v(\bm{x}) \cdot \frac{1}{2}\bm{c} e^{\frac{1}{2}\bm{c}\cdot \bm{x}}.
\end{align*}

Naturally, 
\begin{align*}
\Delta u(\bm{x}) & =e^{\frac{1}{2}\bm{c}\cdot \bm{x}}  \Delta v(\bm{x}) + \frac{1}{2}e^{\frac{1}{2}\bm{c}\cdot \bm{x}}  \bm{c}\cdot \nabla v(\bm{x}) + e^{\frac{1}{2}\bm{c}\cdot \bm{x}}  \frac{1}{2} \bm{c}\cdot \nabla v(\bm{x})+ \frac{1}{4}|\bm{c}|^{2} e^{\frac{1}{2}\bm{c}\cdot \bm{x}}\\
& = e^{\frac{1}{2}\bm{c}\cdot \bm{x}}  \Delta v(\bm{x}) + e^{\frac{1}{2}\bm{c}\cdot \bm{x}}  \bm{c}\cdot \nabla v(\bm{x}) + \frac{1}{4}|\bm{c}|^{2} e^{\frac{1}{2}\bm{c}\cdot \bm{x}},
\end{align*}
and
\begin{align*}
    0=\Delta u -\bm{c}\cdot \nabla u &= e^{\frac{1}{2}\bm{c}\cdot \bm{x}}  \Delta v(\bm{x}) + e^{\frac{1}{2}\bm{c}\cdot \bm{x}}  \bm{c}\cdot \nabla v(\bm{x}) + \frac{1}{4}|\bm{c}|^{2} e^{\frac{1}{2}\bm{c}\cdot \bm{x}}\\
    &- \bm{c} \cdot (e^{\frac{1}{2}\bm{c}\cdot \bm{x}}  \nabla v(\bm{x})+ v(\bm{x}) \cdot \frac{1}{2}\bm{c} e^{\frac{1}{2}\bm{c}\cdot \bm{x}})\\
    &=e^{\frac{1}{2}\bm{c}\cdot \bm{x}}(\Delta v(\bm{x})- \frac{1}{4}|\bm{c}|^{2}v(\bm{x})).
\end{align*}

Hence, we can get 
\begin{align}\label{AB1}
   \Delta v(\bm{x})- \frac{1}{4}|\bm{c}|^{2}v(\bm{x}) =0, 
\end{align}
and let $k^{2}=\frac{1}{4}|\bm{c}|^{2}$. Since the equation \eqref{AB1} is radial symmetric, we can suppose that $v(\bm{x})=w(r), r=|\bm{x}|$,
\begin{align*}
    \Delta v(\bm{x})=w''(r)+\dfrac{2}{r}w'(r),
\end{align*}
then
\begin{align}\label{AB2}
w''(r)+\dfrac{2}{r}w'(r)-k^{2}w(r)=0.
\end{align}

At this point, we let 
\begin{align*}
    w(r)=e^{kr}\overline{w}(r),\quad k=\pm \dfrac{1}{2}|\bm{c}|,
\end{align*}
then we can get 
\begin{align*}
    w'(r)&= e^{kr}(k\overline{w}(r)+\overline{w}'(r)),\\
    w''(r)&= e^{kr}(\overline{w}''(r)+2k\overline{w}'(r)+k^{2}\overline{w}(r)).
\end{align*}

Furthermore, put these into equation \eqref{AB2}, then
\begin{align*}
 0=w''(r)+\dfrac{2}{r}w'(r)-k^{2}w(r)=e^{kr}(\overline{w}''(r)+2k\overline{w}'(r)+\dfrac{2}{r}\overline{w}'(r)+\dfrac{2k}{r}\overline{w}(r)).   
\end{align*}

 In other words,
\begin{align}\label{AB3}
\overline{w}''(r)+2k\overline{w}'(r)+\dfrac{2}{r}\overline{w}'(r)+\dfrac{2k}{r}\overline{w}(r)=0.
\end{align}

Finally, we can obtain the analytical solution of equation \eqref{AB3} using the method of constant variation.
\begin{align*}
    \overline{w}(r)=\dfrac{s}{r},
\end{align*}
where $s$ is a constant. Hence,
\begin{align*}
    v(r)&=e^{kr}\dfrac{s}{r},\\
    u(\bm{x})&=\dfrac{s}{|\bm{x}|}e^{\frac{1}{2}(\bm{c}\cdot \bm{x}-|\bm{c}||\bm{x}|)}.
\end{align*}

Since the fundamental solution of the generalized equation is related to the constant $\bm{c}$, we denote these fundamental solutions as $\Phi_{\bm{c}}(\bm{x})$.

Due to the natural degradation of $\Phi_{\bm{c}}(\bm{x})$ to $\Phi(\bm{x})$ when $\bm{c}=(0,0,0)^{T}$, hence, we can get $s=\frac{1}{4\pi}$, in other words, 
\begin{align*}
    \Phi_{\bm{c}}(\bm{x})=\dfrac{1}{4\pi|\bm{x}|}e^{\frac{1}{2}(\bm{c}\cdot \bm{x}-|\bm{c}||\bm{x}|)}.
\end{align*}

$\hfill \blacksquare$

\section{ derivation of Anisotropic Gauss Formula}\label{C}
This is the proof process of Theorem \ref{THGGL}.

We will prove the theorem in three different situations $\bm{x} \in \Omega$, $\bm{x} \in \partial \Omega$ and $ \bm{x} \in \overline{\Omega}^{c}$, where $\overline{\Omega}^{c}$ represents the complement of the closure of $\Omega$.

First, for $ \bm{x} \in \overline{\Omega}^{c}$, by using the divergence theorem and the fact that $\Phi_{\bm{c}} $ is smooth for $\bm{y} \in \Omega$, $\bm{x} \in \overline{\Omega}^{c}$,
\begin{align*}
   &\int_{\partial \Omega}K_{\bm{c}}(\bm{x}-\bm{y})\cdot \bm{n}(\bm{y}) \d S(\bm{y})  \\
   &=\int_{\partial \Omega} (\nabla \Phi_{\bm{c}}-\bm{c} \cdot  \Phi_{\bm{c}})(\bm{x}-\bm{y})\cdot \bm{n}(\bm{y}) \d S(\bm{y})\\
   &=\int_{\Omega} (\Delta \Phi_{\bm{c}}- \bm{c} \cdot \nabla \Phi_{\bm{c}})  \d \bm{y}\\
   &=\int_{\Omega} 0 \ \d \bm{y} =0.
\end{align*}

In other words, $\forall \bm{c}$, $\forall \bm{x} \in \overline{\Omega}^{c}$,
\begin{align*}
\int_{\partial \Omega}K_{\bm{c}}(\bm{x}-\bm{y})\cdot \bm{n}(\bm{y}) \d S(\bm{y}) = 0.   
\end{align*}

Secondly, for $\bm{x} \in \Omega $, $\Phi_{\bm{c}}(\bm{x}-\bm{y}) $ is not smooth for all $\bm{x} \in \Omega $. In order to overcome this problem, we fix $\varepsilon >0 $ sufficiently small such that $B(\bm{x};\varepsilon)$ which is the ball with $\bm{x}$ as the center and $\varepsilon >0 $ as the radius is contained within $\Omega$. Then
on the region $\Omega-B(\bm{x};\varepsilon)$, $\Phi_{\bm{c}}(\bm{x}-\bm{y}) $ is smooth and we can obtain that
\begin{align*}
    0&=\int_{\Omega-B(\bm{x};\varepsilon)} \Delta \Phi_{\bm{c}}- \bm{c} \cdot \nabla \Phi_{\bm{c}} \d \bm{y} \\
    &=\int_{\partial \Omega - \partial B(\bm{x};\varepsilon)}  (\nabla \Phi_{\bm{c}}-\bm{c} \cdot  \Phi_{\bm{c}})(\bm{x}-\bm{y}) \cdot \bm{n}(\bm{y}) \d S(\bm{y})  \\
    &=\int_{\partial \Omega} (\nabla \Phi_{\bm{c}}-\bm{c} \cdot  \Phi_{\bm{c}})(\bm{x}-\bm{y}) \cdot \bm{n}(\bm{y}) \d S(\bm{y}) \\
    &+\int_{\partial B(\bm{x};\varepsilon)}  (\nabla \Phi_{\bm{c}}-\bm{c} \cdot  \Phi_{\bm{c}})(\bm{x}-\bm{y})\cdot \bm{n}_{B}(\bm{y}) \d S(\bm{y}), 
\end{align*}
where $\bm{n}_{B}$ is the outer unit normal to $B(\bm{x};\varepsilon)$, which is given by 
\begin{align*}
\bm{n}_{B}(\bm{y})=\dfrac{\bm{x}-\bm{y}}{|\bm{x}-\bm{y}|}\triangleq \dfrac{\bm{r}}{|\bm{r}|},
\end{align*}
for $\bm{y} \in \partial B(\bm{x};\varepsilon)$. Denote 
\begin{align*}
F(\bm{c},\varepsilon)=\int_{\partial B(\bm{x};\varepsilon)}  (\nabla \Phi_{\bm{c}}-\bm{c} \cdot  \Phi_{\bm{c}})(\bm{x}-\bm{y})\cdot \bm{n}_{B}(\bm{y}) \d S(\bm{y}). \end{align*}

On the one hand $\forall \bm{c}$,
\begin{align*}
    F(\bm{c},\varepsilon)&= \int_{\partial B(\bm{x};\varepsilon)} \dfrac{1}{4\pi|\bm{r}|}e^{1/2(\bm{c}\cdot \bm{r}-|\bm{c}||\bm{r}|)}(-\dfrac{\bm{r}}{|\bm{r}|^{2}}-\dfrac{1}{2}\bm{c}-\dfrac{1}{2}|\bm{c}|\dfrac{\bm{r}}{|\bm{r}|}) \cdot \dfrac{\bm{r}} {|\bm{r}|} \d S(\bm{y})\\
    &= \int_{\partial B(\bm{x};\varepsilon)} \dfrac{1}{4\pi|\bm{r}|}e^{1/2(\bm{c}\cdot \bm{r}-|\bm{c}||\bm{r}|)}(-\dfrac{1}{|\bm{r}|}-\dfrac{\bm{c}\cdot \bm{r}}{2|\bm{r}|}-\dfrac{1}{2}|\bm{c}|) \d S(\bm{y})\\
    & \leqslant \int_{\partial B(\bm{x};\varepsilon)} \dfrac{1}{4\pi|\bm{r}|}e^{1/2(-|\bm{c}||\bm{r}|-|\bm{c}||\bm{r}|)}(-\dfrac{1}{|\bm{r}|}-\dfrac{\bm{c}\cdot \bm{r}}{2|\bm{r}|}-\dfrac{1}{2}|\bm{c}|) \d S(\bm{y})\\
    &=  \int_{\partial B(\bm{x};\varepsilon)} \dfrac{1}{4\pi|\bm{r}|}e^{(-|\bm{c}||\bm{r}|)}(-\dfrac{1}{|\bm{r}|}-\dfrac{\bm{c}\cdot \bm{r}}{2|\bm{r}|}-\dfrac{1}{2}|\bm{c}|) \d S(\bm{y})\\
    &\leqslant \int_{\partial B(\bm{x};\varepsilon)} \dfrac{1}{4\pi|\bm{r}|}e^{(-|\bm{c}||\bm{r}|)}(-\dfrac{1}{|\bm{r}|}+\dfrac{|\bm{c}||\bm{r}|}{2|\bm{r}|}-\dfrac{1}{2}|\bm{c}|) \d S(\bm{y})\\
    &\leqslant \int_{\partial B(\bm{x};\varepsilon)} \dfrac{1}{4\pi|\bm{r}|}e^{(-|\bm{c}||\bm{r}|)}(-\dfrac{1}{|\bm{r}|}) \d S(\bm{y})\\
    &=  \dfrac{1}{4\pi|\bm{r}|}e^{(-|\bm{c}||\bm{r}|)}(-\dfrac{1}{|\bm{r}|}) \int_{\partial B(\bm{x};\varepsilon)} \d S(\bm{y})\\
    &=-e^{(-|\bm{c}||\bm{r}|)} =-e^{(-|\bm{c}|\varepsilon)}\triangleq F_{1} (\bm{c},\varepsilon).
\end{align*}

On the other hand $\forall \bm{c}$,
\begin{align*}
F(\varepsilon) &\geqslant \int_{\partial B(\bm{x};\varepsilon)} \dfrac{1}{4\pi|\bm{r}|}e^{1/2(|\bm{c}||\bm{r}|-|\bm{c}||\bm{r}|)}(-\dfrac{1}{|\bm{r}|}-\dfrac{\bm{c}\cdot \bm{r}}{2|\bm{r}|}-\dfrac{1}{2}|\bm{c}|) \d S(\bm{y})\\
&=\int_{\partial B(\bm{x};\varepsilon)} \dfrac{1}{4\pi|\bm{r}|}(-\dfrac{1}{|\bm{r}|}-\dfrac{\bm{c}\cdot \bm{r}}{2|\bm{r}|}-\dfrac{1}{2}|\bm{c}|) \d S(\bm{y})\\
&\geqslant \int_{\partial B(\bm{x};\varepsilon)} \dfrac{1}{4\pi|\bm{r}|}(-\dfrac{1}{|\bm{r}|}-\dfrac{|\bm{c}||\bm{r}|}{2|\bm{r}|}-\dfrac{1}{2}|\bm{c}|) \d S(\bm{y})\\
&= \dfrac{1}{4\pi|\bm{r}|}(-\dfrac{1}{|\bm{r}|}-|\bm{c}|) \int_{\partial B(\bm{x};\varepsilon)} \d S(\bm{y})\\
&=-1-|\bm{c}||\bm{r}|=-1-|\bm{c}|\varepsilon \triangleq F_{2} (\bm{c},\varepsilon).
\end{align*}

Letting the fixed $\varepsilon \to 0^{+}$, then we can find that 
\begin{align*}
    &\lim\limits_{\varepsilon \to 0^{+}} F_{1}(\bm{c},\varepsilon)=-1, \quad \lim\limits_{\varepsilon \to 0^{+}} F_{2}(\bm{c},\varepsilon)=-1.
\end{align*}

Therefore, it can be inferred from the squeeze theorem that
\begin{align*}
   \lim\limits_{\varepsilon \to 0^{+}}F(\bm{c},\varepsilon)=-1.
\end{align*}

In other words, $\forall \bm{c}$, $\forall \bm{x} \in  \Omega$,
\begin{align*}
 \int_{\partial \Omega}K_{\bm{c}}(\bm{x}-\bm{y})\cdot \bm{n}(\bm{y}) \d S(\bm{y})=1.   
\end{align*}

Last, we consider the case $\bm{x} \in \partial \Omega$. In this case, $\Phi_{\bm{c}}(\bm{x}-\bm{y})$ is not defined at $\bm{y}=\bm{x}$.
Fix $\bm{x} \in \partial \Omega$. Let $B(\bm{x};\varepsilon)$ be the ball of radius $r$ about $\bm{x}$. Let $\Omega_{\varepsilon} \equiv \Omega-(\Omega \cap B(\bm{x};\varepsilon))$, and $\mathcal{C}_{\varepsilon} \equiv \{ \bm{y} \in \partial B(\bm{x};\varepsilon):\bm{n}\cdot \bm{y} <0 \}$, and $ \widetilde{\mathcal{C}_{\varepsilon}} \equiv \partial \Omega_{\varepsilon} \cap \mathcal{C}_{\varepsilon}$.

Firstly, we provide a lemma that needs to be used  and provide the proof of it.
\begin{lemma}\label{CC}
    For $\widetilde{\mathcal{C}_{\varepsilon}}$ and $\mathcal{C}_{\varepsilon}$ as defined above, we have
    \begin{align*}
        \int_{\widetilde{\mathcal{C}_{\varepsilon}}} \d S(\bm{y})=\int_{\mathcal{C}_{\varepsilon}} \d S(\bm{y}) +O(\varepsilon^{3}). 
    \end{align*}
\end{lemma}

    We just need to show that the surface area of $\widetilde{\mathcal{C}_{\varepsilon}}-\mathcal{C}_{\varepsilon}$ is $O(\varepsilon^{3})$. The surface area is
approximately the surface area of the base times the height. Now the surface area of the
base is $O(\varepsilon)$. Therefore, we just need to show that the height is $O(\varepsilon^{2})$.

Without loss of generality, we let $\bm{x} = \bm{0}$. Now, by assumption, $\partial \Omega$ is $C^{2}$. Therefore, $\partial \Omega$ can be written as the graph of a $C^{2}$ function $f$: $\mathbb{R}^{2}  \to \mathbb{R}$ such that $f(\bm{0}) = 0$ and $\nabla f (\bm{0})$. Therefore, if $\bm{y}=(y_{1},y_{2},y_{3}) \in \mathcal{C}_{\varepsilon} - \widetilde{\mathcal{C}_{\varepsilon}}$, then
\begin{align*}
    |y_{3}|\leqslant|f(y_{1},y_{2})|\leqslant C|(y_{1},y_{2})|^{2}\leqslant C \varepsilon^{2},
\end{align*}
by using Taylor’s theorem. Therefore, the height is $O(\varepsilon^{2})$ and the lemma follows.

Then, we start to prove this situation. First, we note that
\begin{align*}
      0&=\int_{\Omega_{\varepsilon}} \Delta \Phi_{\bm{c}}- \bm{c} \cdot \nabla \Phi_{\bm{c}} \d \bm{y} \\
      &=\int_{\partial \Omega_{\varepsilon}} (\nabla \Phi_{\bm{c}}-\bm{c} \cdot  \Phi_{\bm{c}})(\bm{x}-\bm{y}) \cdot \bm{n}(\bm{y}) \d S(\bm{y})  \\
      &=\int_{\partial \Omega_{\varepsilon}-\widetilde{\mathcal{C}_{\varepsilon}}}(\nabla \Phi_{\bm{c}}-\bm{c} \cdot  \Phi_{\bm{c}})(\bm{x}-\bm{y}) \cdot \bm{n}(\bm{y}) \d S(\bm{y}) \\
      &+ \int_{\widetilde{\mathcal{C}_{\varepsilon}}}(\nabla \Phi_{\bm{c}}-\bm{c} \cdot  \Phi_{\bm{c}})(\bm{x}-\bm{y}) \cdot \bm{n}(\bm{y}) \d S(\bm{y}),
\end{align*}
where $\bm{n}$ is the outer unit normal to $\Omega_{\varepsilon}$, and
\begin{align*}
\bm{n}(\bm{y})=\dfrac{\bm{x}-\bm{y}}{|\bm{x}-\bm{y}|}\triangleq \dfrac{\bm{r}}{|\bm{r}|}.
\end{align*}

Hence, on the one hand,  
\begin{align*}
F_{\partial \Omega} (\bm{c},\varepsilon) \triangleq &\int_{\widetilde{\mathcal{C}_{\varepsilon}}}(\nabla \Phi_{\bm{c}}-\bm{c} \cdot  \Phi_{\bm{c}})(\bm{x}-\bm{y}) \cdot \bm{n}(\bm{y}) \d S(\bm{y})\\
      &= \int_{\widetilde{\mathcal{C}_{\varepsilon}}} \dfrac{1}{4\pi|\bm{r}|}e^{1/2(\bm{c}\cdot \bm{r}-|\bm{c}||\bm{r}|)}(-\dfrac{1}{|\bm{r}|}-\dfrac{\bm{c}\cdot \bm{r}}{2|\bm{r}|}-\dfrac{1}{2}|\bm{c}|) \d S(\bm{y}) \\  
      &\leqslant \int_{\widetilde{\mathcal{C}_{\varepsilon}}} \dfrac{1}{4\pi|\bm{r}|}e^{(-|\bm{c}||\bm{r}|)}(-\dfrac{1}{|\bm{r}|}) \d S(\bm{y})\\
      &=\dfrac{1}{4\pi|\bm{r}|}e^{(-|\bm{c}||\bm{r}|)}(-\dfrac{1}{|\bm{r}|})\int_{\widetilde{\mathcal{C}_{\varepsilon}}} \d S(\bm{y}) \\
      &=\dfrac{1}{4\pi \varepsilon}e^{(-|\bm{c}|\varepsilon)}(-\dfrac{1}{\varepsilon})\int_{\widetilde{\mathcal{C}_{\varepsilon}}} \d S(\bm{y})\\
      &= \dfrac{1}{4\pi \varepsilon}e^{(-|\bm{c}|\varepsilon)}(-\dfrac{1}{\varepsilon})\cdot (\int_{\mathcal{C}_{\varepsilon}} \d S(\bm{y}) + O(\varepsilon^{3})) \quad (\text{by Lemma \ref{CC}})\\
      &=-\dfrac{e^{(-|\bm{c}|\varepsilon)}}{2}+O(\varepsilon) \triangleq F_{3} (\bm{c},\varepsilon),
\end{align*}
where the expansion and contraction process of the omitted techniques of inequalities is the same as the calculation of $F_{1} (\bm{c},\varepsilon)$. On the other hand,
\begin{align*}
 F_{\partial \Omega} (\bm{c},\varepsilon) &\geqslant   \dfrac{1}{4\pi|\bm{r}|}(-\dfrac{1}{|\bm{r}|}-|\bm{c}|) \int_{\widetilde{\mathcal{C}_{\varepsilon}}} \d S(\bm{y})\\
      &=\dfrac{1}{4\pi \varepsilon}(-\dfrac{1}{ \varepsilon}-|\bm{c}|) \int_{\widetilde{\mathcal{C}_{\varepsilon}}} \d S(\bm{y})\\
      &=\dfrac{1}{4\pi \varepsilon}(-\dfrac{1}{ \varepsilon}-|\bm{c}|) \cdot (\int_{\mathcal{C}_{\varepsilon}} \d S(\bm{y})+O(\varepsilon^{3}))  \quad (\text{by Lemma \ref{CC}})\\
      &=-\dfrac{1}{2}-\dfrac{|\bm{c}|\varepsilon}{2}+ O(\varepsilon) \triangleq F_{4} (\bm{c},\varepsilon).  
\end{align*}

Taking the limit as $\varepsilon \to 0^{+}$, we have
\begin{align*}
    &\lim\limits_{\varepsilon \to 0^{+}} F_{3}(\bm{c},\varepsilon)=-\dfrac{1}{2}, \quad \lim\limits_{\varepsilon \to 0^{+}} F_{4}(\bm{c},\varepsilon)=-\dfrac{1}{2}.
\end{align*}

Therefore, it can be inferred from the squeeze theorem that
\begin{align*}
   \lim\limits_{\varepsilon \to 0^{+}}F_{\partial \Omega}(\bm{c},\varepsilon)=-\dfrac{1}{2}.
\end{align*}

In other words, $\forall \bm{c}$, $\forall \bm{x} \in \partial \Omega$,
\begin{align*}
\int_{\partial \Omega}K_{\bm{c}}(\bm{x}-\bm{y})\cdot \bm{n}(\bm{y}) \d S(\bm{y}) = \dfrac{1}{2}.   
\end{align*}
$\hfill \blacksquare$




\end{document}